\documentclass[aps,pra,twocolumn,notitlepage]{revtex4-2}
\usepackage{amsmath,amssymb}	
\usepackage{mathtools}
\usepackage{natbib}
\usepackage{bm} 
\usepackage{braket}
\usepackage{hyperref}
\usepackage{url}
\usepackage{graphicx}
\usepackage{color}
\usepackage{xcolor}
\usepackage{array}
\renewcommand{\Im}{\text{Im}}
\renewcommand{\Re}{\text{Re}}
\newcommand{\HO}{\hat{\mathcal{H}}}
\newcommand{\LO}{\hat{\mathcal{L}}}

\newcommand{\GO}{\hat{\mathcal{G}}}
\newcommand{\AO}{\hat{\mathcal{A}}}
\newcommand{\BO}{\hat{\mathcal{B}}}
\newcommand{\CO}{\hat{\mathcal{C}}}
\newcommand{\OO}{\hat{\mathcal{O}}}
\newcommand{\IO}{\hat{\mathcal{I}}}

\newcommand{\TO}{\hat{\mathcal{T}}}
\begin{document}
\title{Time-dependent Schrieffer-Wolff-Lindblad Perturbation Theory: measurement-induced dephasing and second-order Stark shift in dispersive readout}
\date{\today}
\author{Moein Malekakhlagh}
\email{Electronic address: moein.malekakhlagh@ibm.com}
\author{Easwar Magesan}
\email{Electronic address: emagesa@us.ibm.com}
\author{Luke C. G. Govia}
\email{Electronic address: lcggovia@ibm.com}
\affiliation{IBM Quantum, IBM Thomas J. Watson Research Center, 1101 Kitchawan Rd, Yorktown Heights, NY , USA, 10598}
\begin{abstract}
We develop a time-dependent Schrieffer-Wolff-Lindblad perturbation theory to study effective interactions for driven open quantum systems. The starting point of our analysis is a given Lindblad equation, based on which we obtain an effective (averaged) map that describes the renormalization of both the Hamiltonian and collapse operators due to the drive. As a case study, we apply this method to the dispersive readout of a transmon qubit and derive an effective disperive map that describes measurement-induced dephasing and Stark shift for the transmon. The effective map we derive is completely positive and trace-preserving under adiabatic resonator response. To benchmark our method, we demonstrate good agreement with a numerical computation of the effective rates via the Lindbladian spectrum. Our results are also in agreement with, and extend upon, an earlier derivation of such effects by Gambetta et al. \cite{Gambetta_Qubit-photon_2006} (\href{https://journals.aps.org/pra/abstract/10.1103/PhysRevA.74.042318}{Phys. Rev. A 74, 042318}) using the positive P-representation for the resonator field.  
\end{abstract}

\maketitle
%%%%%%%%%%%%%%%%%%%%%%%%%%%%% Sec:Intro %%%%%%%%%%%%%%%%%%%%%%%%%%%%%%%%%%%%
\section{Introduction}
\label{Sec:Intro}

The Lindblad master equation \cite{Lindblad_Generators_1976, Gorini_Completely_1976} describes the evolution of an open quantum system \cite{Breuer_Theory_2002, Gardiner_Quantum_2004} undergoing Markovian dynamics and constitutes a generalization of the unitary description of a closed system. Theoretical methods for characterizing the Lindblad dynamics \cite{Albert_Symmetries_2014, Albert_Geometry_2016, Manzano_Short_2020} of a system are crucial for understanding error processes and system behavior in the presence of environmental effects. In the case of unitary evolution, there are a wide variety of time-independent and -dependent perturbative methods, for example Rayleigh-Schr\"{o}dinger Perturbation Theory (RSPT) \cite{Rayleigh_Theory_1896, Schrodinger_Quantisierung_1926, Sakurai_Modern_1995, Griffiths_Introduction_2016}, Schrieffer-Wolff Perturbation Theory (SWPT) \cite{Schrieffer_Relation_1966, Soliverez_General_1981, Boissonneault_Dispersive_2009, Bravyi_Schrieffer_2011, Gambetta_Analytic_2011, Malekakhlagh_Lifetime_2020, Petrescu_Lifetime_2020, Magesan_Effective_2020, Malekakhlagh_First-Principles_2020, Xiao_Perturbative_2021, Petrescu_Accurate_2021, Malekakhlagh_Mitigating_2022, Malekakhlagh_Optimization_2022}, Magnus expansion \cite{Magnus_Exponential_1954, Blanes_Magnus_2009, Blanes_Pedagogical_2010}, Dyson series \cite{Dyson_Radiation_1949, Shillito_Fast_2021}, and Multi-Scale Perturbation Theory (MSPT) \cite{Bender_Multiple_1996, Bender_Advanced_1999}.

Perturbation theories for open quantum systems fall into two broad categories based on the treatment of system-environment interaction: (i) accounting for the interaction as a perturbation, and deriving effective master equations \cite{Grifoni_Driven_1998, Nathan_Universal_2020, Petrescu_Lifetime_2020}, or (ii) adopting a master equation as the starting point and computing effective interactions \cite{Li_Perturbative_2014, Azouit_Adiabatic_2016, Villegas_Application_2016, Shishkov_Perturbation_2020, Hanai_Intrinsic_2021}. Approach (i) provides a more precise description of the environment as one can relate the effective relaxation rates to the environment spectral function. However, the perturbative expansion of this approach can suffer from divergences near system-drive resonances as the zeroth-order system self-energy vanishes, due to the fact that it is purely real and has no contribution from the dissipation. 
Approach (ii) resolves this divergence by regulating the effective interactions through a non-zero relaxation rate in the starting model. The limitation, however, is that it cannot account for possible corrections to dissipation rates due to the sensitivity of the environment spectral function on the dynamic (Stark) shifts of system energies due to Hamiltonian interactions or drive. Here, we develop a perturbation theory based on approach (ii).                

In this paper, we extend SWPT \cite{Schrieffer_Relation_1966, Soliverez_General_1981, Boissonneault_Dispersive_2009, Bravyi_Schrieffer_2011, Gambetta_Analytic_2011, Malekakhlagh_Lifetime_2020, Petrescu_Lifetime_2020, Magesan_Effective_2020, Malekakhlagh_First-Principles_2020, Xiao_Perturbative_2021, Petrescu_Accurate_2021, Malekakhlagh_Mitigating_2022, Malekakhlagh_Optimization_2022} to the level of the Lindblad master equation, a method that we coin Schrieffer-Wolff-Lindblad Perturbation Theory (SWLPT). Using SWLPT, we are able to study effective interactions for a driven open quantum system. This includes the renormalization of Hamiltonian parameters due to the interplay between drive and dissipation, as well as the renormalization or emergence of incoherent mechanisms. Our development of SWLPT is time-dependent which accounts for corrections caused by transients in the drive pulse. Applying SWLPT to the dispersive readout of a transmon  qubit \cite{Koch_Charge_2007}, we derive an effective dispersive map modeling the low-power behavior of measurement-induced dephasing and Stark shift. We demonstrate how such perturbative calculations can be performed in terms of oscillator (bosonic) modes, which agrees with and extends the earlier studies that employed two-level descriptions of the qubit \cite{Gambetta_Qubit-photon_2006, Boissonneault_Dispersive_2009}. Interestingly, under adiabatic response, we find the effective map to be representable in Lindblad form, and hence Completely Positive and Trace-Preserving (CPTP) \cite{Nielsen_Quantum_2002}.              

The remainder of this paper is organized as follows. Section~\ref{Sec:SWLPT} summarizes the generalization of the SWPT method into SWLPT for studying Lindblad dynamics. In Sec.~\ref{Sec:Model}, we introduce an approximate dispersive model for the measurement of a transmon qubit. Section~\ref{Sec:EffMap} provides an effective dispersive map for the system evolution, derived using SWLPT, describing effective Stark shifts and dephasing rates. In Sec.~\ref{Sec:MeasIndStSh}, we compare and contrast our results with a previous derivation in Ref.~\cite{Gambetta_Qubit-photon_2006}, as well as with a numerical computation of the effective rates. In Sec.~\ref{Sec:Transient}, we show the application of SWLPT for studying the transient dynamics of effective interactions. Section~\ref{Sec:Summary} provides the summary and future directions.     

The paper is supplemented with seven appendices. Appendix~\ref{App:Vector} reviews a method for vectorization of Lindblad dynamics in terms of an extended Schr\"odinger-like equation following Ref.~\cite{Yi_Effective_2001}. In Appendix~\ref{App:DispTrans}, we discuss a displacement transformation on the resonator mode and the corresponding coherent mean-field response. The main results of the paper are derived in Appendix~\ref{App:SWLPT}, where we obtain an effective dispersive map for readout using the generalized SWLPT method. In Appendix~\ref{App:EffMapProp}, we summarize properties of the effective adiabatic dispersive map. In Appendix~\ref{App:LindForm}, we show that the effective adiabatic dispersive map can be expressed in Lindblad form. In Appendix~\ref{App:Transient}, we analyze the time-dependent nature of the perturbation and transient dynamics. Appendix~\ref{App:EigState} provides perturbative solutions for instantaneous measurement eigenstates.             
%%%%%%%%%%%%%%%%%%%%%%%%%%%%%%%%%%%%%%%%%%%%%%%%%%%%%%%%%%%%%%%%%%%%%%%%%%%%%%%%%

%%%%%%%%%%%%%%%%%%%%%%%%%%% Sec: SWLPT %%%%%%%%%%%%%%%%%%%%%%%%%%%%%%%%%%%%%%%%%
\section{Schrieffer-Wolff-Lindblad Perturbation Theory}
\label{Sec:SWLPT}
Our development of a time-dependent SWLPT formalism is based on combining two methods: (i) vectorization and representation of a given Lindbladian in terms of an extended Hamiltonian \cite{Yi_Effective_2001, Prosen_Quantization_2010} (Appendix~\ref{App:Vector}), and (ii) generalization of the time-dependent SWPT that was developed at the Hamiltonian level \cite{Schrieffer_Relation_1966, Soliverez_General_1981, Boissonneault_Dispersive_2009, Bravyi_Schrieffer_2011, Gambetta_Analytic_2011, Malekakhlagh_Lifetime_2020, Petrescu_Lifetime_2020, Magesan_Effective_2020, Malekakhlagh_First-Principles_2020, Xiao_Perturbative_2021, Petrescu_Accurate_2021, Malekakhlagh_Mitigating_2022, Malekakhlagh_Optimization_2022}. 

Given a Lindblad equation 
\begin{align}
\partial_t\hat{\rho}(t)=-i[\hat{H}_s+\hat{H}_d(t),\hat{\rho}(t)]+\sum\limits_{j}\gamma_j \mathcal{D}	[\hat{C}_j]\hat{\rho}(t) \;,
\label{eqn:SWLPT-Pure Lindblad Eq}
\end{align}
with $\hat{H}_s$, $\hat{H}_d(t)$, $\gamma_j \mathcal{D}[\hat{C}_j]$ as the system Hamiltonian, drive Hamiltonian, and the $j^{th}$ dissipator, respectively, we are looking for a Schr\"odinger-like equation
\begin{align}
\partial_t \ket{\Psi_{\hat{\rho}}(t)} = -i \HO_u (t) \ket{\Psi_{\hat{\rho}}(t)}  \;,  
\label{eqn:SWLPT-Pure Schro-like Eq}
\end{align}
where $\HO_u(t)$ is the extended Hamiltonian and $\ket{\Psi_{\hat{\rho}}(t)}$ is the vectorized density matrix. We follow the vectorization introduced in Ref.~\cite{Yi_Effective_2001}. Assuming the solution $\hat{\rho}(t)=\sum_{mn} \rho_{mn}(t)\ket{m}\bra{n}$ is  expressed in an orthonormal basis $\{\ket{n}\}$, the corresponding vectorized solution is $\ket{\Psi_{\hat{\rho}}(t)}=\sum_{mn} \rho_{mn}(t) \ket{m_l}\ket{n_r}$. Here, $l$ and $r$ denote the states of the original (left) and auxiliary (right) copies of the Hilbert space. Consequently, one finds the extended Hamiltonian $\HO_u(t)$ as
\begin{align}
& \HO_u(t) \equiv \HO_l(t)-\HO_r(t) + \HO_{\gamma} \;, 
\label{eqn-SWLPT:Pure-Def of Hu}\\
&\HO_{\gamma} \equiv 	\sum\limits_j i \gamma_j \left( \CO_{j,l}\CO_{j,r}-\frac{1}{2}\CO_{j,l}^{\dag} \CO_{j,l} -\frac{1}{2} \CO_{j,r}^{\dag}\CO_{j,r} \right) \;,
\label{eqn-SWLPT:Pure-Def of H_gamma}
\end{align}
where $\HO_{l/r}(t)$ denote independent left and right copies of the system and drive Hamiltonian, and $\CO_{j,l/r}$ denote the left and right extended collapse operators. Moreover, the left and right extended operators corresponding to an arbitrary operator $\hat{O}$ are defined as $\OO_{l}\equiv \hat{O}\otimes \hat{I}$ and $\OO_{r}\equiv \hat{I} \otimes \hat{O}^*$, respectively (Appendix~\ref{App:Vector}).

Given the extended Hamiltonian $\HO_u(t)$, we apply a time-dependent SW transformation, the details of which depends on the hierarchy of system and drive energy scales, as well as the quantities we wish to compute. Based on such a hierarchy, we define bare and interaction extended Hamiltonians $\HO_u(t)\equiv \HO_0 + \HO_{\text{int}}(t)$, and work in the interaction frame with respect to $\HO_0$ such that $\HO_I(t)\equiv \TO_0^{-1}(t)\HO_{\text{int}}(t)\TO_0(t)$, where $\TO_0(t)\equiv \exp(-i\HO_0 t)$. 

The effective extended Hamiltonian is defined by a similarity SW transformation 
\begin{align}
\HO_{I,\text{eff}}(t)\equiv \TO_{\text{SW}}^{-1}(t)[\HO_I(t)-i\partial_t]\TO_{\text{SW}}(t) \;,
\label{eqn:SWLPT-SWPT Trans}
\end{align}
where $\TO_{\text{SW}}(t)\equiv \exp [-i \GO(t)]$ and $\GO(t)$ is the generator. One minor distinction of SWLPT with respect to SWPT is that $\GO(t)$ is not necessarily Hermitian, and thus $\TO_{\text{SW}}(t)$ not unitary. However, due to the similar form of the operator transformations, this does not alter the final SWLPT equations compared to SWPT. In particular, writing $\GO(t)$ and $\HO_{I,\text{eff}}(t)$ as power series in $\HO_I(t)$ and using the Baker-Campbell-Hausdorff (BCH) lemma \cite{Baker_Alternants_1905, Campbell_Law_1896, Hausdorff_Symbolische_1906} results in a set of perturbative operator-valued ODEs for $\GO_n(t)$ and corresponding solutions for $\HO_{I,\text{eff}}^{(n)}(t)$ for $n\geq 1$ (see Appendix C of Ref.~\cite{Malekakhlagh_First-Principles_2020} for derivation). 

Defining $\mathcal{S}(\bullet)$ and $\mathcal{N}(\bullet)$ as projections onto an effective subspace (which we choose) and its compliment, respectively, we find the first-order SWLPT equation as
\begin{subequations}
\begin{align}
&\HO_{I,\text{eff}}^{(1)}(t)=\mathcal{S} \Big(\HO_{I}(t)\Big) \;,
\label{eqn:SWLPT-Def of H_I,eff^(1)}\\
&\dot{\GO}_{1}(t)=\mathcal{N} \Big(\HO_{I}(t)\Big) \;,
\label{eqn:SWLPT-Def of G_1}
\end{align}
\end{subequations}
the second order as 
\begin{subequations}
\begin{align}
&\HO_{I,\text{eff}}^{(2)}(t)=\mathcal{S} \Big(i[\GO_1(t),\HO_{I}(t)]-\frac{i}{2} [\GO_1(t),\dot{\GO}_1(t)] \Big) \;,
\label{eqn:SWLPT-Def of H_I,eff^(2)}\\
&\dot{\GO}_{2}(t)=\mathcal{N} \Big(i[\GO_1(t),\HO_{I}(t)]-\frac{i}{2} [\GO_1(t),\dot{\GO}_1(t)] \Big) \;,
\label{eqn:SWLPT-Def of G_2}
\end{align}
\end{subequations}
and the third order as
\begin{subequations}
\begin{align}
\begin{split}
\hat{\mathcal{H}}_{\text{I,eff}}^{(3)}(t)&=\mathcal{S}\Big(-\frac{i}{2}[\hat{\mathcal{G}}_1(t),\dot{\hat{\mathcal{G}}}_2(t)]-\frac{i}{2}[\hat{\mathcal{G}}_2(t),\dot{\hat{\mathcal{G}}}_1(t)]\\
&+\frac{1}{6}[\hat{\mathcal{G}}_1(t),[\hat{\mathcal{G}}_1(t),\dot{\hat{\mathcal{G}}}_1(t)]]+i[\hat{\mathcal{G}}_2(t),\hat{\mathcal{H}}_I(t)]\\
&-\frac{1}{2}[\hat{\mathcal{G}}_1(t),[\hat{\mathcal{G}}_1(t),\hat{\mathcal{H}}_I(t)]] \Big)\;,
\end{split}
\label{eqn:SWLPT-Def of H_I,eff^(3)}\\
\begin{split}
\dot{\hat{\mathcal{G}}}_3(t)&=\mathcal{N}\Big(-\frac{i}{2}[\hat{\mathcal{G}}_1(t),\dot{\hat{\mathcal{G}}}_2(t)]-\frac{i}{2}[\hat{\mathcal{G}}_2(t),\dot{\hat{\mathcal{G}}}_1(t)]\\
&+\frac{1}{6}[\hat{\mathcal{G}}_1(t),[\hat{\mathcal{G}}_1(t),\dot{\hat{\mathcal{G}}}_1(t)]]+i[\hat{\mathcal{G}}_2(t),\hat{\mathcal{H}}_I(t)]\\
&-\frac{1}{2}[\hat{\mathcal{G}}_1(t),[\hat{\mathcal{G}}_1(t),\hat{\mathcal{H}}_I(t)]] \Big)\;.
\end{split}
\label{eqn:SWLPT-Def of G_3}
\end{align}
\end{subequations}
Two common scenarios for the choice of $\mathcal{S}(\bullet)$  are diagonalization \cite{Xiao_Perturbative_2021, Malekakhlagh_Optimization_2022} and block-diagonaliation \cite{Magesan_Effective_2020, Malekakhlagh_First-Principles_2020, Malekakhlagh_Mitigating_2022}. Moreover, the natural choice for setting the initial condition of the operator-valued ODEs~(\ref{eqn:SWLPT-Def of G_1}),~(\ref{eqn:SWLPT-Def of G_2}) and~~(\ref{eqn:SWLPT-Def of G_3}) is to pick $\GO_n(t)$ based on the particular solution, i.e. indefinite integration over the driven contributions of the right-hand side.

Note that there is flexibility in the definition of $\HO_0$ and $\HO_{\text{int}}(t)$. We conjecture the choice of $\HO_0$ determines whether $\HO_{I,\text{eff}}(t)$ is expressible in Lindblad form at arbitrary truncation order. In particular, we face the choices of (i) including $\HO_{\gamma}$ of Eq.~(\ref{eqn-SWLPT:Pure-Def of H_gamma}) in $\HO_0$, or (ii) breaking and keeping the diagonal terms $\sum_j (-i/2) \gamma_j (\CO_{j,l}^{\dag} \CO_{j,l} + \CO_{j,r}^{\dag}\CO_{j,r})$ in $\HO_0$ and the collapse terms $\sum_j i \gamma_j \CO_{j,l} \CO_{j,r}$ in $\HO_{\text{int}}(t)$. Following (i), given that $\HO_{\gamma}$ is quadratic, $\HO_0$ can in principle be exactly diagonalized using a symplectic (Bogoliubov) transformation, also referred to as the third quantization \cite{Prosen_Quantization_2010, Mcdonald_Exact_2022}. Here, since both $\HO_0$ and $\HO_{\text{int}}(t)$ are initially in a Lindblad form, it may be possible that SWLPT Eqs.~(\ref{eqn:SWLPT-Def of H_I,eff^(1)})--(\ref{eqn:SWLPT-Def of G_3}) conserve the Lindblad forms for $\GO_n(t)$ and $\HO_{I,\text{eff}}^{(n)}(t)$ at arbitrary order. Understanding what, if any, further conditions need to be satisfied for this to hold is an interesting direction for future work.

The pre-SWLPT symplectic transformation, however, is a challenging computation on its own, especially for a multimode system, and could lead to a more complex form of the Hamiltonian interaction in $\HO_{\text{int}}(t)$. Following (ii), diagonalization of the collapse terms is postponed to the third order in SWPLT. However, since neither $\HO_0$ nor $\HO_{\text{int}}(t)$ are in Lindblad forms to begin with, $\HO_{I,\text{eff}}^{(n)}(t)$ is not in general expressible in a Lindblad form either. In this work, we follow choice (ii) to simplify the perturbative calculations. Importantly, we find that under adiabatic response \footnote{For a time-dependent $\HO_{\text{int}}(t)$ of the form $\HO_{\text{int}}(t) = f(t)\HO_{\text{int}}$, then adiabatic response implies that we only keep terms proportional to $f(t)$ in our solutions, and drop all terms proportional to its derivatives, as these are assumed to be vanishingly small.} and up to the third-order SWLPT it is still possible to re-express the effective map for dispersive readout in a Lindblad form. Thus, while it does not in general guarantee Lindblad form, in specific cases the simpler approach (ii) can still be used to derive faithful descriptions of the relevant physics.

%%%%%%%%%%%%%%%%%%%% Fig: Dispersive measurement Schematic %%%%%%%%%%%%%%%%%%
\begin{figure}[t!]
\centering
\includegraphics[scale=0.38]{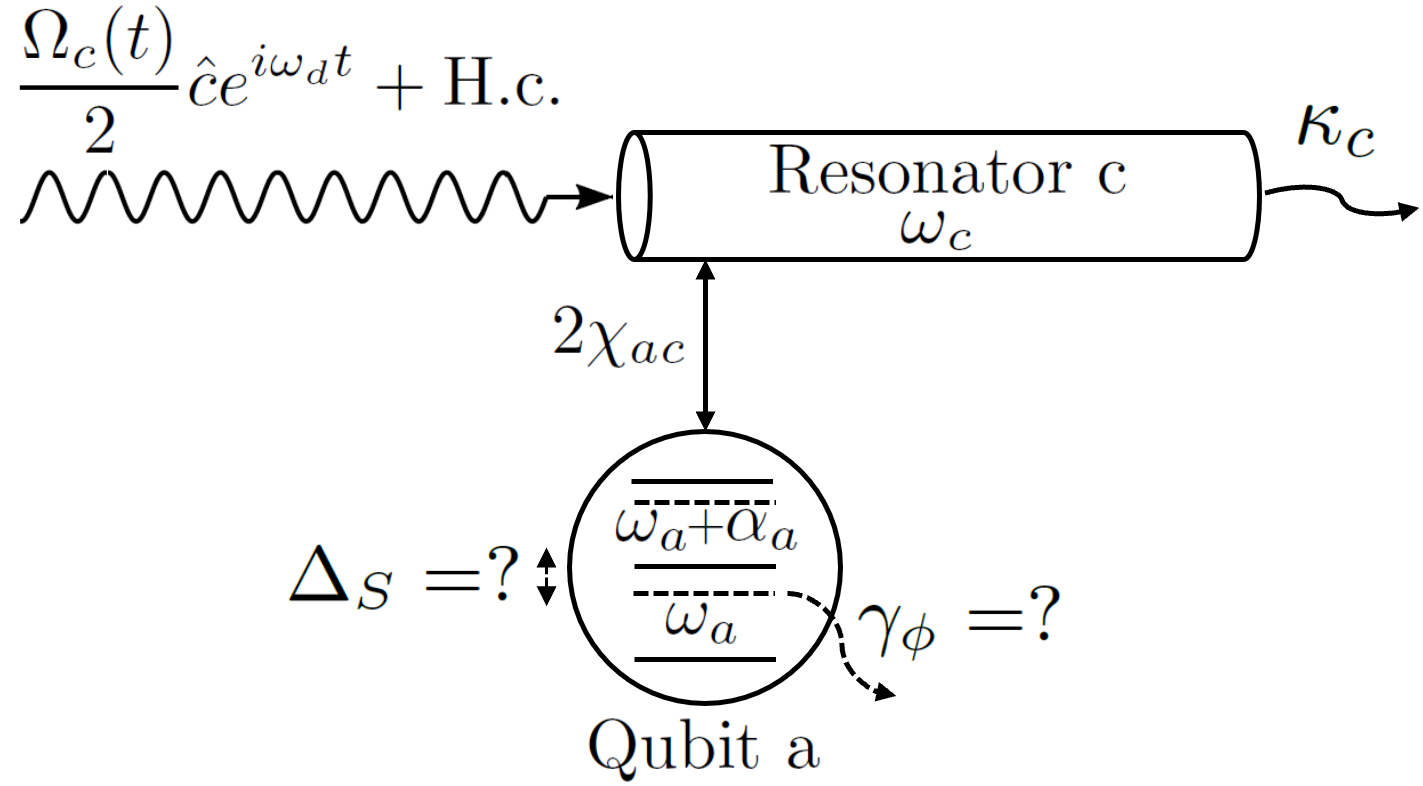}
\caption{Schematic of the measurement setup based on a dispersive Kerr model described in Eqs.~(\ref{eqn:Model-Def of Hs})--(\ref{eqn:Model-DispKerrLindblad}). Features of the bare model and effective map are shown with solid and dashed lines, respectively. The measurement tone causes a Stark shift, $\Delta_S$, and measurement-induced dephasing, with rate $\gamma_{\phi}$, on the qubit, for which we developed the SWLPT method starting with a Lindblad evolution for the system.}
\label{fig:DispMeasSchmeatic}
\end{figure}
%%%%%%%%%%%%%%%%%%%%%%%%%%%%%%%%%%%%%%%%%%%%%%%%%%%%%%%%%%%%%%%%%%%%%%%%%%%%%%   

%%%%%%%%%%%%%%%%%%%%%%%%%%%%%%%%%%%%%%%%%%%%%%%%%%%%%%%%%%%%%%%%%%%%%%%%%%%%%%%%

%%%%%%%%%%%%%%%%%%%%%%%%%%%%% Sec:Model %%%%%%%%%%%%%%%%%%%%%%%%%%%%%%%%%%%%
\section{Model for dispersive readout}
\label{Sec:Model}
	 
We consider a transmon qubit \cite{Koch_Charge_2007}, modeled as a nonlinear quantum Kerr oscillator, coupled dispersively to a driven dissipative resonator mode (see Fig.~\ref{fig:DispMeasSchmeatic}). In the rotating frame of the drive, and under the Rotating-Wave Approximation (RWA), the system and drive Hamiltonian read 
\begin{align} 
\begin{split}
\hat{H}_s &\equiv \Delta_{ad} \hat{a}^{\dag}  \hat{a}+\frac{1}{2} \alpha_a \hat{a}^{\dag}\hat{a}^{\dag}\hat{a}\hat{a}\\
&+ \Delta_{cd} \hat{c}^{\dag}\hat{c}+2\chi_{ac}\hat{a}^{\dag}\hat{a}\hat{c}^{\dag}\hat{c} \;,
\end{split}
\label{eqn:Model-Def of Hs}\\
\hat{H}_d(t) &\equiv \frac{\Omega_c(t)}{2}\left(\hat{c}+\hat{c}^{\dag}\right)	 \;,
\label{eqn:Model-Def of Hd}
\end{align}
where for the normal mode $j\in\{a,c\}$ ($a$ - transmon, $c$ - resonator), $\Delta_{jd}\equiv \omega_j-\omega_d$ is the detuning from drive, $\alpha_a$ is the qubit anharmonicity, $2\chi_{ac}$ is the full dispersive shift, and $\Omega_c(t)$ is the time-dependent measurement pulse on the resonator mode.   

Furthermore, to model resonator relaxation, we include the dissipator $\kappa_c\mathcal{D}[\hat{c}]$ in a Lindblad equation for the system denisty matrix
\begin{align}
\partial_t \hat{\rho}(t)= - i \left[\hat{H}_s+\hat{H}_d(t),\hat{\rho}(t)\right] + \kappa_c \mathcal{D}[\hat{c}]\hat{\rho}(t) \;,
\label{eqn:Model-DispKerrLindblad}
\end{align} 
where $\mathcal{D}[\hat{c}]\hat{\rho}(t)\equiv \hat{c}\hat{\rho}(t)\hat{c}^{\dag}-(1/2)\{\hat{c}^{\dag}\hat{c},\hat{\rho}(t)\}$ and $\kappa_c=\kappa_{c,\downarrow}$ is the downward relaxation rate. Given the fact that at thermal equilibrium $\kappa_{c,\uparrow}/\kappa_{c,\downarrow}\approx\exp(-\hbar\omega_c/K_B T)$, and assuming a state-of-the-art cryogenic temperature of $T \approx~$10--20 mK \footnote{See Bluefors products: \href{https://bluefors.com/products/xld-dilution-refrigerator}{https://bluefors.com/products/xld-dilution-refrigerator}}, the upward rate $\kappa_{c,\uparrow}$ is negligible. 

A few remarks are in order regarding adopting the dispersive Kerr Hamiltonian~(\ref{eqn:Model-Def of Hs}) as our starting model. First, in writing Eq.~(\ref{eqn:Model-Def of Hs}), we assume in principle that the underlying physical interaction is a transverse qubit-resonator coupling of the form $-g_{ac} (\hat{a}-\hat{a}^{\dag})(\hat{c}-\hat{c}^{\dag})$. Up to the leading order in the dispersive limit, i.e.~when $|\Delta_{ac}|\gg g_{ac}$, this results in a number-number Kerr interaction with $\chi_{ac} = [\alpha_a/(\Delta_{ac}+\alpha_c)](g_{ac}^2/\Delta_{ac})$ \cite{Koch_Charge_2007, Boissonneault_Dispersive_2009}. 

Second, the transverse interaction can however lead to a variety of non-QND (off-diagonal with respect to the normal qubit mode) interaction forms which can be computed using normal mode expansion techniques \cite{Nigg_BlackBox_2012, Minev_EPR_2020, Malekakhlagh_Optimization_2022}. Such non-QND contributions have been shown to result in a renormalization of the qubit energy relaxation rate \cite{Malekakhlagh_Lifetime_2020, Petrescu_Lifetime_2020, Hanai_Intrinsic_2021}, as well as leakage to high-excitation qubit states \cite{Sank_Measurement-Induced_2016, Malekakhlagh_Optimization_2022, Shillito_Dynamics_2022}. To focus on the main mechanism behind measurement-induced dephasing, and not complicate our analytical formulation, we work with the dispersive Kerr model. For this reason, we also do not include the resonator-induced Purcell decay of the transmon qubit \cite{Houck_Controlling_2008, Malekakhlagh_Cutoff-Free_2017, Scheer_Computational_2018} in our model, which to lowest order would add a term of the form $(g_{ac}/\Delta_{ac})^2\kappa_c \mathcal{D}[\hat{a}]\hat{\rho}(t)$ to Eq.~\eqref{eqn:Model-DispKerrLindblad}.

Third, under a two-level approximation for the qubit, our dispersive Kerr model is in principle equivalent to earlier studies based on a dispersive JC formulation \cite{Gambetta_Qubit-photon_2006, Boissonneault_Dispersive_2009}. The dispersive JC interaction form $\chi_{ac}\hat{\sigma}^z\hat{c}^{\dag}\hat{c}$, however, results in the resonator-frequency offset $\chi_{ac}\hat{c}^{\dag}\hat{c}$, compared to the dispersive Kerr interaction $2\chi_{ac}\hat{a}^{\dag}\hat{a}\hat{c}^{\dag}\hat{c}$, which needs to be accounted for in any comparison between the two models (Sec.~\ref{Sec:MeasIndStSh}).

%%%%%%%%%%%%%%%%%%%%%%%%%%%%% Sec: Effective map %%%%%%%%%%%%%%%%%%%%%%%%%%%%%%%
\section{Effective dispersive map for readout}
\label{Sec:EffMap}

We next apply the SWLPT method of Sec.~\ref{Sec:SWLPT} on the dispersive readout model of Sec.~\ref{Sec:Model} and derive an effective dipserive map that captures both the induced Stark shift and measurement-induced dephasing. Moreover, we show that under adiabatic response, the effective map is CPTP, and so in the low-power regime, the map can be considered as a valid quantum channel.

Applying the described vectorization method to Eqs.~(\ref{eqn:Model-Def of Hs})--(\ref{eqn:Model-DispKerrLindblad}), the corresponding extended Hamiltonian reads 
$\HO_u(t) \equiv \HO_l (t) - \HO_r (t) +\HO_{\kappa}$, where $\HO_l(t)$ and $\HO_r (t)$ are independent copies of the system and drive Hamiltonian
\begin{align}
\begin{split}
\HO_j (t) &\equiv \Delta_{ad} \hat{a}_j^{\dag}  \hat{a}_j+\frac{1}{2} \alpha_a \hat{a}_j^{\dag}\hat{a}_j^{\dag}\hat{a}_j\hat{a}_j\\
&+ \Delta_{cd} \hat{c}_j^{\dag}\hat{c}_j+2\chi_{ac}\hat{a}_j^{\dag}\hat{a}_j\hat{c}_j^{\dag}\hat{c}_j \\
&+ \frac{\Omega_c(t)}{2}\left(\hat{c}_j+\hat{c}_j^{\dag}\right)	 \;, \quad j=l,r \;,
\end{split}
\label{eqn:EffMap-Def of H_l&H_r}
\end{align}
and $\HO_{\kappa}$ is the corresponding representation of the resonator relaxation as  
\begin{align}
\HO_{\kappa} \equiv i \kappa_c \left( \hat{c}_l\hat{c}_r -\frac{1}{2}\hat{c}_l^{\dag} \hat{c}_l - \frac{1}{2} \hat{c}_r^{\dag}\hat{c}_r\right)  \;.
\label{eqn:EffMap-Def of H_kappa}
\end{align}
In such a representation, the left and right modes in Eqs.~(\ref{eqn:EffMap-Def of H_l&H_r})--(\ref{eqn:EffMap-Def of H_kappa}) obey the standard bosonic commutation relations $[\hat{a}_j,\hat{a}_k^{\dag}]=[\hat{c}_j,\hat{c}_k^{\dag}]=\delta_{jk}\IO $, with all other commutators being zero, i.e. $[\hat{a}_j,\hat{a}_k]=[\hat{c}_j,\hat{c}_k]=[\hat{a}_j,\hat{c}_k]=[\hat{a}_j,\hat{c}_k^{\dag}]=[\hat{a}_j^{\dag},\hat{c}_k]=[\hat{a}_j^{\dag},\hat{c}_k^{\dag}]=0$, for $j$, $k$ $\in \{l,r\}$.   

We then compute an \textit{effective} extended Hamiltonian by diagonalizing $\HO_u(t)$ as $\HO_{I,\text{eff}}(t) \equiv \TO_{\text{diag}}^{-1}(t) [\HO_{u}(t)-i\partial_t]\TO_{\text{diag}}(t)$. The diagonalization transformation $\TO_{\text{diag}}(t)$ is comprised of three transformations: $\TO_{\text{diag}}(t) \equiv \TO_D[\eta_c(t)]
\TO_{0}(t) \TO_{\text{SW}}(t)$. First, $\TO_D[\eta_c(t)]$ is a coherent displacement of the resonator modes $\hat{c}_l$ and $\hat{c}_r$, by $\eta_c(t)$ and $\eta_c^*(t)$, respectively, where $\eta_c(t)$ is the classical resonator response whose evolution is described by the equation (Appendix~\ref{App:DispTrans}):    
\begin{align}
\dot{\eta}_c (t) + \left(i\Delta_{cd}+\frac{\kappa_c}{2}\right)\eta_c(t)=-\frac{i}{2}\Omega_c(t) \;.
\label{eqn:EffMap-Def of eta_c(t)}
\end{align}
The steady-state photon number is found as $n_{c}\equiv |\eta_{c,\text{ss}}|^2 = (\Omega_c/2)^2/[\Delta_{cd}^2+(\kappa_c/2)^2]$. We note that it is more common to define qubit-state-dependent $\eta_c(t)$ and photon number (Refs.~\cite{Gambetta_Qubit-photon_2006, Boissonneault_Dispersive_2009} and table~\ref{tab:CompWithGambetta}). In our theory, however, such a dependence is accounted for in the SW expansion. Second, $\TO_{0}(t)$ is the transformation to the interaction frame with respect to the undriven diagonal part of Eqs.~(\ref{eqn:EffMap-Def of H_l&H_r})--(\ref{eqn:EffMap-Def of H_kappa}), generated by $\HO_0$ defined in Eqs.~\eqref{Eq:SWLPT-Def of H_l0} and \eqref{Eq:SWLPT-Def of H_r0}. Third, $\TO_{\text{SW}}(t)\equiv \exp[-i\GO (t)]$ is a generalized time-dependent SW transformation as described in Sec.~\ref{Sec:SWLPT}.

Implementing SWLPT up to third order, we arrive at the following effective extended Hamiltonian under adiabatic resonator response, where we keep only terms proportional to $\eta_c(t)$ and drop its derivatives (Appendix~\ref{App:SWLPT}):
\begin{align}
\begin{split}
\HO_{I,\text{eff}}^{\text{ad}}(t)  = &\:\:\:\:\:\: 2\chi_{ac}|\eta_c(t)|^2 \hat{n}_{al}-2\chi_{ac}|\eta_c(t)|^2 \hat{n}_{ar} \\
&-\frac{4\chi_{ac}^2|\eta_c(t)|^2}{\hat{\Delta}_{cdl}} \hat{n}_{al}^2 +\frac{4\chi_{ac}^2|\eta_c(t)|^2}{\hat{\Delta}_{cdr}} \hat{n}_{ar}^2 \\ 
&+i \frac{4\chi_{ac}^2 \kappa_c |\eta_c(t)|^2}{\hat{\Delta}_{cdl} \hat{\Delta}_{cdr}} \hat{n}_{al} \hat{n}_{ar} \;,\\ 
\end{split}
\label{eqn:EffMap-Adiab H_I,eff}
\end{align}
where $\hat{n}_{al}\equiv \hat{a}_l^{\dag}\hat{a}_l$ and $\hat{n}_{ar}\equiv \hat{a}_r^{\dag}\hat{a}_r$ are the left and right qubit number operators in the effective (diagonal) frame, and $\hat{\Delta}_{cdl}$ and $\hat{\Delta}_{cdr}$ are qubit-state-dependent resonator-drive detunings defined as     
\begin{subequations}
\begin{align}
&\hat{\Delta}_{cdl} \equiv \Delta_{cd}-i\frac{\kappa_c}{2}+2\chi_{ac}\hat{n}_{al}\;,
\label{eqn:EffMap-Def of Delta_cdl}\\
&\hat{\Delta}_{cdr} \equiv \Delta_{cd}+i\frac{\kappa_c}{2}+2\chi_{ac}\hat{n}_{ar} \;.
\label{eqn:EffMap-Def of Delta_cdr}
\end{align}
\end{subequations}
Equation~(\ref{eqn:EffMap-Adiab H_I,eff}) is one of the main results of this paper, which gives the effective dynamics for the qubit degrees of freedom. $\HO_{I,\text{eff}}^{\text{ad}}(t)$ provides the renormalization of qubit transition frequencies and the accompanying dephasing rates as a function of the measurement drive for the multi-level anharmonic oscillator description of the transmon qubit, which extends the two-level descriptions previously studied \cite{Gambetta_Qubit-photon_2006, Boissonneault_Dispersive_2009}. See Sec.~\ref{Sec:Transient} and Appendix~\ref{App:Transient} for the full time-dependent form.

SWLPT also provides a means to perturbatively compute the instantaneous eigenstates (Appendix~\ref{App:EigState}). The effective and initial frames are related via the diagonalization transformation $\TO_{\text{diag}}(t)$ such that $\ket{\Psi_{\hat{\rho}}(t)} = \TO_{\text{diag}}(t) \ket{\Psi_{\hat{\rho},\text{eff}}(t)}$. This mapping makes the role of resonator degrees of freedom more explicit. For example, the effective-frame eigenstate with labels $n_{al}=1$ and $n_{ar}=0$ can be expressed in the initial frame up to first order as (Appendix~\ref{App:EigState})    
\begin{align}
\begin{split}
\left\{1-\frac{2\chi_{ac}\eta_c(t)\left[\hat{c}_l^{\dag}-\eta_c^*(t)\right]}{\Delta_{cd}-i\frac{\kappa_c}{2}+2\chi_{ac}}\right\}\ket{1_{al},\eta_c(t)}\ket{0_{ar},\eta_c^*(t)} \;.
\end{split}
\label{eqn:EffMap-LabFrameRep of 1_L0_R}
\end{align}
Therefore, up to zeroth order, the left and right resonator modes are in the coherent states $\eta_c(t)$ and $\eta_c^*(t)$, respectively. However, there also exist higher-order qubit-dependent resonator excitations on top of the coherent states, implying drive-induced interaction between the normal qubit and resonator modes.     

Effective time evolution under $\HO_{I,\text{eff}}^{\text{ad}}(t)$, i.e. $\exp[-i\int_{0}^{t}dt' \HO_{I,\text{eff}}^{\text{ad}}(t')]$, referred to as the effective adiabatic dispersive map, holds desirable properties (Appendices~\ref{App:EffMapProp} and~\ref{App:LindForm}). In particular, the spectrum of $\HO_{I,\text{eff}}^{\text{ad}}(t)$ obeys:    
\begin{subequations}
\begin{align}
& E_{n_{a},n_{a}}^{\text{ad}}(t) = 0 \;,
\label{eqn:EffMap-E_nn=0}\\
& E_{n_{al},n_{ar}}^{\text{ad}}(t) = - E_{n_{ar},n_{al}}^{\text{ad}*}(t) \;,
\label{eqn:EffMap-E_mn=-E_nm^*}\\
& \Im\{E_{n_{al},n_{ar}}^{\text{ad}}(t)\}<0 \;, 
\label{eqn:EffMap-Im(E_mn)<0}
\end{align}
\end{subequations}
where $\HO_{I,\text{eff}}^{\text{ad}}(t)\ket{n_{al}}\ket{n_{ar}}\equiv E_{n_{al},n_{ar}}^{\text{ad}}(t)\ket{n_{al}}\ket{n_{ar}}$ (Appendix~\ref{SubApp:EffDispSpec}). Equations~(\ref{eqn:EffMap-E_nn=0})--(\ref{eqn:EffMap-Im(E_mn)<0}) imply that the map is TP, Hermiticity-Preserving (HP) and contracting, respectively (Appendix~\ref{App:EffMapProp}).

%%%%%%%%%%%%%%%%%%%% Table: Comparison with Gambetta et al %%%%%%%%%%%%%%%%%%%%%
\begin{table*}
\begin{tabular}{|c|c|c|}
\hline
& Gambetta et al. (2006) \cite{Gambetta_Qubit-photon_2006} & This work\\
\hline\hline
Measurement-induced dephasing & $\gamma_{\phi} = \frac{\chi_{ac}^2 \kappa_c}{\Delta_{cd}^2+\chi_{ac}^2+\kappa_c^2/4} \left(n_{c,+}+n_{c,-}\right)$ & $\gamma_{\phi} = \frac{2\chi_{ac}^2\kappa_c}{(\Delta_{cd}+2\chi_{ac})^2+(\kappa_c/2)^2}n_c $\\
\hline\hline
Stark shift& $\Delta_S = 2 \chi_{ac}(P)n_c(P)$ & $\Delta_{S} \approx \left[2\chi_{ac}-\frac{4\chi_{ac}^2(\Delta_{cd}+2\chi_{ac})}{(\Delta_{cd}+2\chi_{ac})^2+(\kappa_c/2)^2}\right]n_c$ \\
\hline\hline
Steady-state photon number& $n_{c,\pm} \equiv |\eta_{c,\pm,\text{ss}}|^2 = \frac{(\Omega_c/2)^2}{(\Delta_{cd}\pm \chi_{ac})^2+(\kappa_c/2)^2}$ & $n_c \equiv |\eta_{c,\text{ss}}|^2 = \frac{(\Omega_c/2)^2}{\Delta_{cd}^2+(\kappa_c/2)^2} $ \\
\hline
\end{tabular}
\caption{Comparison between our effective rates [Eqs.~(\ref{eqn:MeasIndStSh-Delta_S Sol})--(\ref{eqn:MeasIndStSh-gamma_phi Sol})] and the result of Gambetta et al.~\cite{Gambetta_Qubit-photon_2006} [section IV and Eqs.~(5.20)--(5.21)]. The expressions for measurement-induced dephasing are indeed the same. The apparent difference arises from the different model for the transmon in each study, i.e.~anharmonic oscillator versus two-level system. For the Stark shift, however, Gambetta et al.~employed a phenomenological model with nonlinear dependence on the input power $P$ to describe high-power measurements \cite{Gambetta_Qubit-photon_2006}.}
\label{tab:CompWithGambetta}
\end{table*}
%%%%%%%%%%%%%%%%%%%%%%%%%%%%%%%%%%%%%%%%%%%%%%%%%%%%%%%%%%%%%%%%%%%%%%%%%%%%%%%%

Moreover, starting from Eq.~(\ref{eqn:EffMap-Adiab H_I,eff}) and reverting the vectorization, we can show that the effective adiabatic dispersive map takes the following Lindblad form (Appendix~\ref{App:LindForm}): 
\begin{align}
\dot{\hat{\rho}}_{\text{I,eff}}^{\text{ad}}(t)=-i[\hat{H}_{\text{I,eff}}^{\text{ad}}(t),\hat{\rho}_{\text{I,eff}}^{\text{ad}}(t)]+\mathcal{D}[\hat{C}_{\text{I,eff}}^{\text{ad}}(t)]\hat{\rho}_{\text{I,eff}}^{\text{ad}}(t) \;.
\label{eqn:EffMap-Eff Lind dynamics}
\end{align}
The effective Hamiltonian $\hat{H}_{\text{I,eff}}^{\text{ad}}(t)$, which contains the first- and second-order Stark shifts, is given by,
\begin{align}
\begin{split}
\hat{H}_{\text{I,eff}}^{\text{ad}}(t)&=2\chi_{ac}|\eta_c(t)|^2 \hat{n}_{a} \\
&-\frac{4\chi_{ac}^2|\eta_c(t)|^2(\Delta_{cd}+2\chi_{ac}\hat{n}_{a})}{(\Delta_{cd}+2\chi_{ac}\hat{n}_{a})^2+(\kappa_c/2)^2}\hat{n}_{a}^2 \;,
\label{eqn:EffMap-Def of H_I,eff^ad}
\end{split}
\end{align}
while the effective collapse operator $\hat{C}_{I,\text{eff}}^{\text{ad}}(t)$ models measurement-induced dephasing:
\begin{align}
\begin{split}
\hat{C}_{\text{I,eff}}^{\text{ad}}(t) = \frac{\sqrt{4\chi_{ac}^2\kappa_c|\eta_c(t)|^2}\hat{n}_{a}}{\Delta_{cd}-i\kappa_c/2+2\chi_{ac}\hat{n}_a} \;.
\end{split}
\label{eqn:EffMap-Def of C_I,eff^ad}
\end{align} 
Given that the effective adiabatic map has a Lindblad representation, it is guaranteed to be CPTP.  

On the precision of the SWLPT expansion for readout, which leads to Eq.~(\ref{eqn:EffMap-Adiab H_I,eff}), we note that it is in powers of the collective interaction form $2\chi_{ac}\eta_c(t)$, and more reliable when the interaction is smaller than the underlying transition frequency detunings, i.e. $|2\chi_{ac}\eta_c(t)|<|\braket{\hat{\Delta}_{cdj}}|$ for $j\in\{l,r\}$ (Appendix~\ref{App:SWLPT}). Using the steady-state expression for $\eta_c(t)$, and $\hat{\Delta}_{cdj}$ with the qubit in the first excited state, one finds:    
\begin{align}
|\chi_{ac}\Omega_c|<\sqrt{\Delta_{cd}^2+\left(\frac{\kappa_c}{2}\right)^2}\sqrt{(\Delta_{cd}+2\chi_{ac})^2+\left(\frac{\kappa_c}{2}\right)^2} \;.
\label{eqn:EffMap-validity of SWPT}
\end{align}
In summary, the SWLPT expansion is more valid for larger $\Delta_{cd}$ and $\kappa_c$, and weaker $\chi_{ac}$ and $\Omega_c$. For improved readout, however, a common choice is to drive in between the two resonances, i.e. $\Delta_{cd}=-\chi_{ac}$, making condition~(\ref{eqn:EffMap-validity of SWPT}) more stringent as $|\chi_{ac}\Omega_c|<\chi_{ac}^2+(\kappa_c/2)^2$.

%%%%%%%%%%%%%%%%%%%%%%%%%%%%%%%%%%%%%%%%%%%%%%%%%%%%%%%%%%%%%%%%%%%%%%%%%%%%%

%%%%%%%%%%%%%%%%%% Sec:Meas-Ind Dephasing and Stark Shift %%%%%%%%%%%%%%%%%%%
\section{Measurement-induced dephasing and second-order Stark shift}
\label{Sec:MeasIndStSh}

We next provide the leading-order expressions for measurement-induced dephasing and Stark shift, discuss the connection with former studies, and demonstrate good agreement with numerical computation of such rates. 

%%%%%%%%%%%%%% Fig: Effective Rates %%%%%%%%%%%%%%%%%
\begin{figure}[t!]
\centering
\includegraphics[scale=0.133]{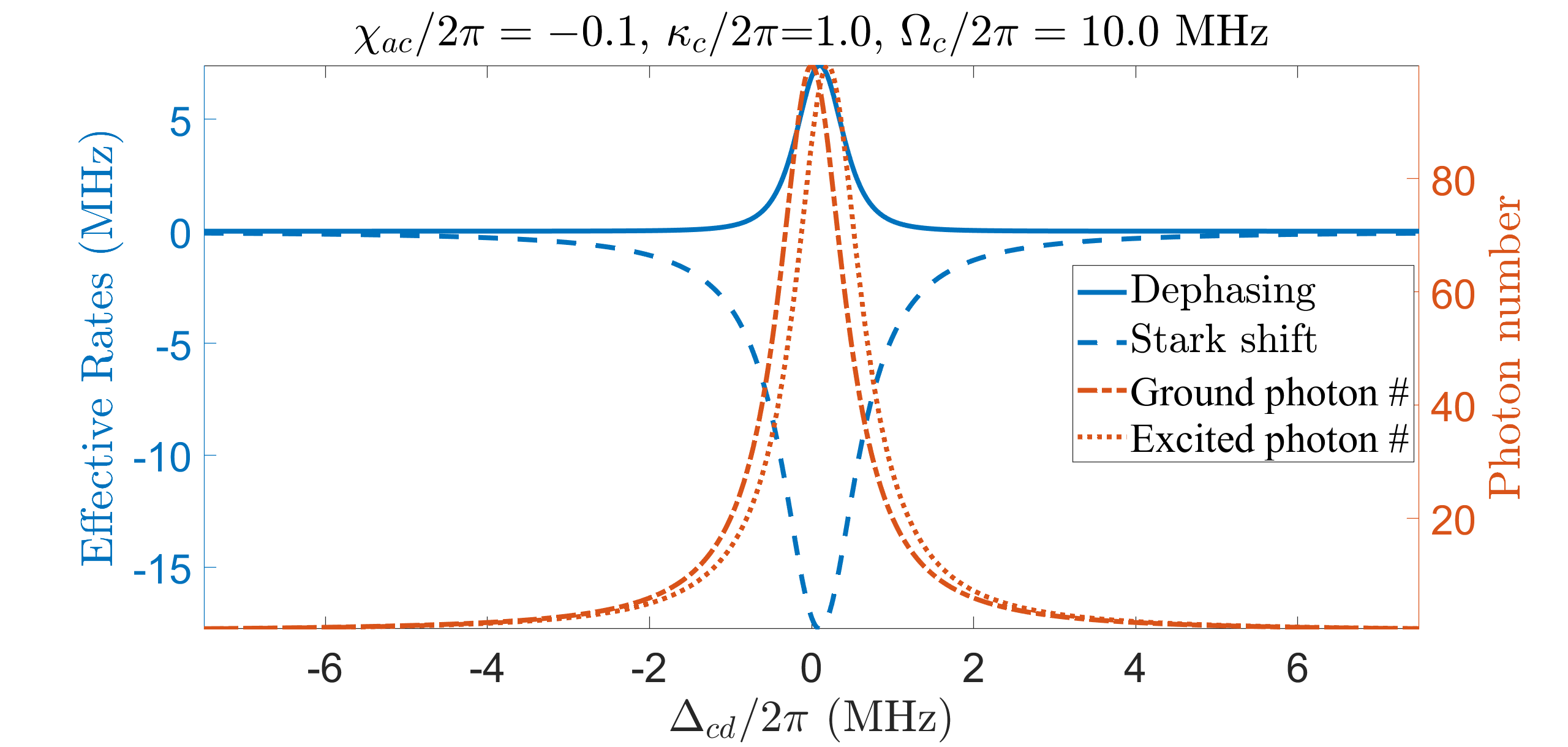}\\
\includegraphics[scale=0.133]{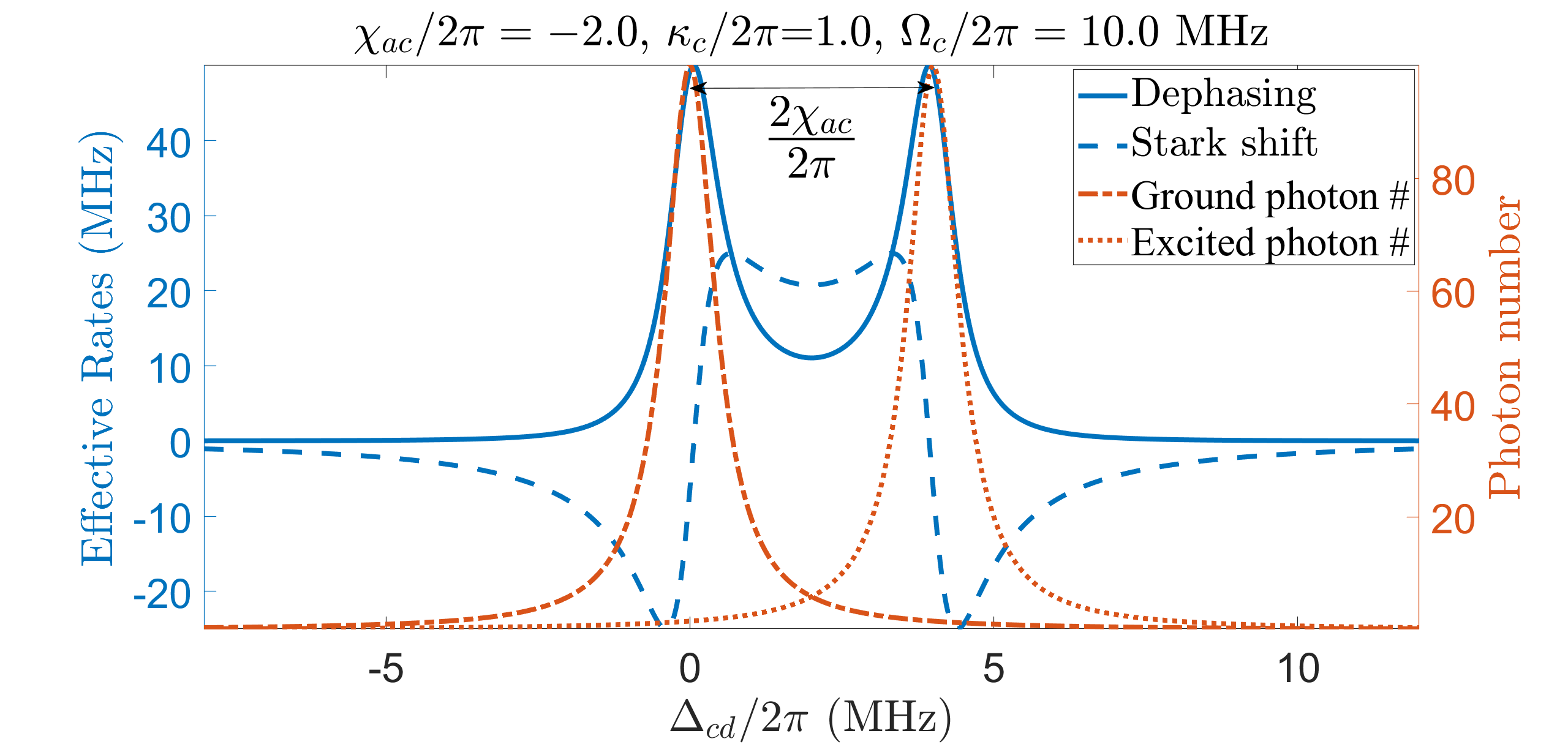}\\
\caption{Measurement-induced dephasing $\gamma_{\phi}$ [Eq.~(\ref{eqn:MeasIndStSh-gamma_phi Sol})] and Stark shift $\Delta_{S}$ [Eq.~(\ref{eqn:MeasIndStSh-Delta_S Sol})], along with qubit-state-dependent resonator photon numbers, as a function of resonator-drive detuning $\Delta_{cd}$. System parameters are set to $\kappa_{c}/2\pi=1$ and $\Omega_c/2\pi=10$ MHz. Panels (a) and (b) correspond to $\chi_{ac}/2\pi$ of $-0.1$ and $-2.0$ MHz, respectively. The ground and excited photon numbers, defined as $(\Omega_c/2)^2/[\Delta_{cd}^2+(\kappa_c/2)^2]$ and $(\Omega_c/2)^2/[(\Delta_{cd}+2\chi_{ac})^2+(\kappa_c/2)^2]$, respectively, are shown for clarity, despite our distinct perturbative expansion that employs explicitly only the ground photon number.}
\label{fig:MeasIndStSh-EffRates}
\end{figure}
%%%%%%%%%%%%%%%%%%%%%%%%%%%%%%%%%%%%%%%%%%%%%%%%%%%%%%%%%%%%%%%%%%%%%%%%%%%%%

Following the vectorized notation, the Stark shift and measurement-induced dephasing are obtained as the real and imaginary parts of the eigenvalue of $\HO_{I,\text{eff}}^{\text{ad}}(t)$ in Eq.~(\ref{eqn:EffMap-Adiab H_I,eff}) corresponding to the state $\ket{1_{al}}\ket{0_{ar}}$, i.e.~$\ket{1_{a}}\bra{0_{a}}$ in density matrix notation, as     
\begin{subequations}
\begin{align}
&\Delta_{S}(t) \approx \left[2\chi_{ac}-\frac{4\chi_{ac}^2(\Delta_{cd}+2\chi_{ac})}{(\Delta_{cd}+2\chi_{ac})^2+(\kappa_c/2)^2}\right]|\eta_c(t)|^2 \;,
\label{eqn:MeasIndStSh-Delta_S Sol}\\
&\gamma_{\phi}(t) \approx \frac{2\chi_{ac}^2\kappa_c}{(\Delta_{cd}+2\chi_{ac})^2+(\kappa_c/2)^2}|\eta_c(t)|^2 \;.
\label{eqn:MeasIndStSh-gamma_phi Sol}
\end{align} 
\end{subequations}
On top of the common 2$\chi_{ac}$-per-photon Stark shift, Eq.~(\ref{eqn:MeasIndStSh-Delta_S Sol}) contains a second-order correction proportional to $\chi_{ac}^2$ [second term of Eq.~(\ref{eqn:MeasIndStSh-Delta_S Sol})]. Such a correction was less noted in the context of dispersive measurement, but studied later on for the Resonator-Induced Phase (RIP) gate \cite{Cross_Optimized_2015, Paik_Experimental_2016, Malekakhlagh_Optimization_2022}. Interestingly, measurement-induced dephasing is of the same origin and order as the second-order Stark shift, where one finds $\gamma_{\phi}(t)=-(1/2)[\kappa_c/(\Delta_{cd}+2\chi_{ac})]\Delta_S^{(2)}(t)$.

Figure~\ref{fig:MeasIndStSh-EffRates} shows $\Delta_{S}$ and $\gamma_{\phi}$ as a function of $\Delta_{cd}$ for different ratios of $|\chi_{ac}|/\kappa_c$ and for time-independent $\Omega_c$. In particular, we observe qualitatively distinct behavior for $2|\chi_{ac}|<\kappa_c$ and $2|\chi_{ac}|>\kappa_c$. For sufficiently small $|\chi_{ac}|/\kappa_c$, both $\gamma_{\phi}$ and $\Delta_S$ demonstrate a single collective peak centered in the middle of the ground and excited  resonances at $\Delta_{cd}=-\chi_{ac}$ [panel (a)]. Enhancing $|\chi_{ac}|/\kappa_c$ results in the splitting of $\Delta_S$ and $\gamma_{\phi}$, and also a positive $\Delta_{S}$ in between the resonances [panel (b)]. 

%%%%%%%%%%%%%% Fig: Comparison with Gambetta et al %%%%%%%%%%%%%%%%%
\begin{figure}[t!]
\centering
\includegraphics[scale=0.138]{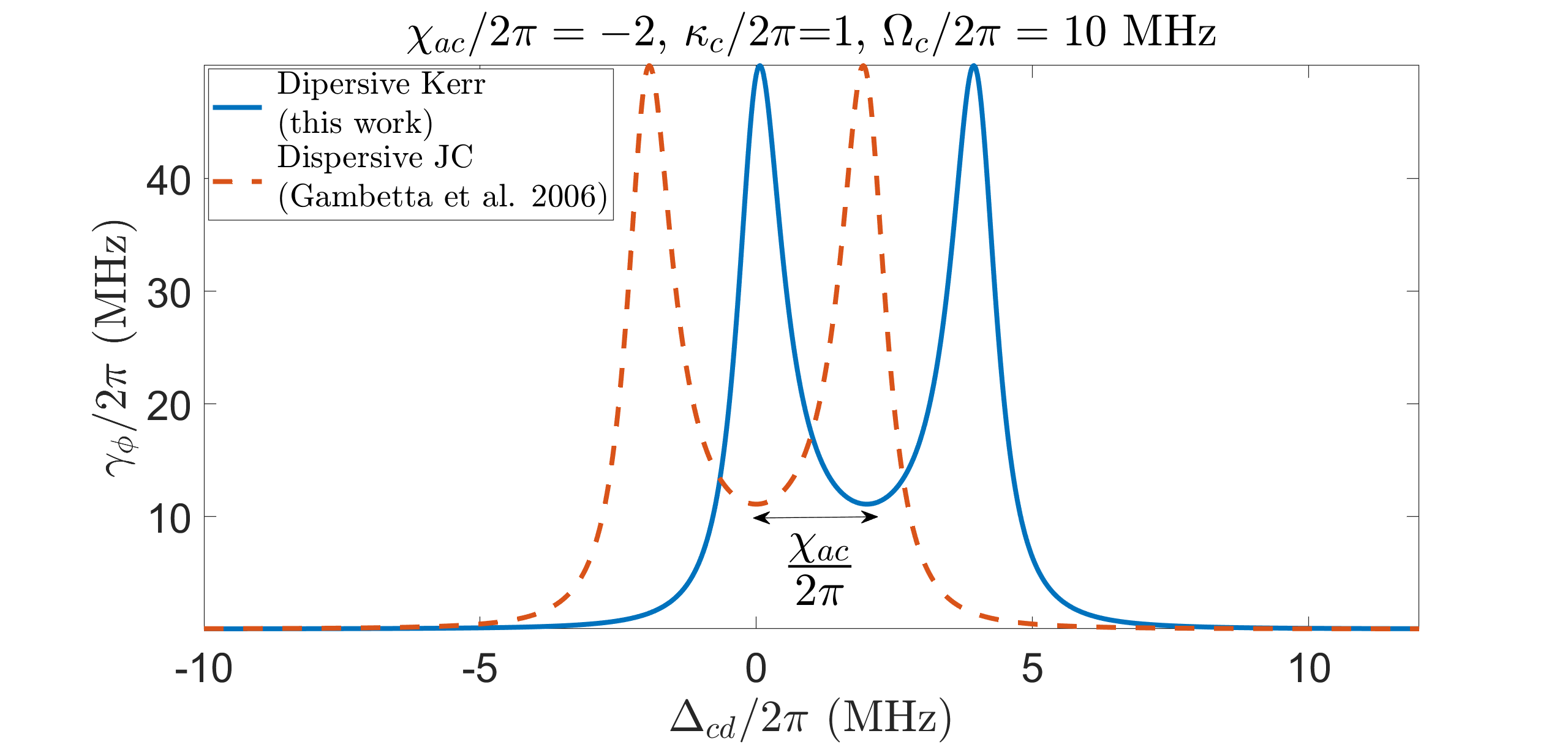}
\caption{Comparison between the expressions for measurement-induced dephasing in this work and Gambetta et al. \cite{Gambetta_Qubit-photon_2006} (see also Table~\ref{tab:CompWithGambetta}). They can be mapped by a $\chi_{ac}$ shift in the definition of $\Delta_{cd}$. System and drive parameters are the same as panel (b) of Fig.~\ref{fig:MeasIndStSh-EffRates}.}
\label{fig:MeasIndStSh-CompWithGambetta}
\end{figure}
%%%%%%%%%%%%%%%%%%%%%%%%%%%%%%%%%%%%%%%%%%%%%%%%%%%%%%%%%%%%%%%%%%%%%%%%%%%%%

Equation~(\ref{eqn:MeasIndStSh-gamma_phi Sol}) for $\gamma_{\phi}$ is in agreement with that of Gambetta et al. \cite{Gambetta_Qubit-photon_2006}. The apparent difference in the expressions is due to the fact that our starting point is the dispersive Kerr model, which models the transmon qubit as an anharmonic oscillator, while Ref.~\cite{Gambetta_Qubit-photon_2006} models the transmon as a two-level system.
% to adopting a dispersive Kerr model as opposed to a dispersive JC model (\color{red}this may be a bit confusing\color{black}), and the two can be mapped 
Our model can be mapped to that of Ref.~\cite{Gambetta_Qubit-photon_2006} by an offset of $\chi_{ac}$ in the definition of $\Delta_{cd}$ as shown in Fig.~\ref{fig:MeasIndStSh-CompWithGambetta}. Moreover, Table~\ref{tab:CompWithGambetta} provides a detailed comparison. 

\begin{figure}[t!]
\centering
\includegraphics[scale=0.335]{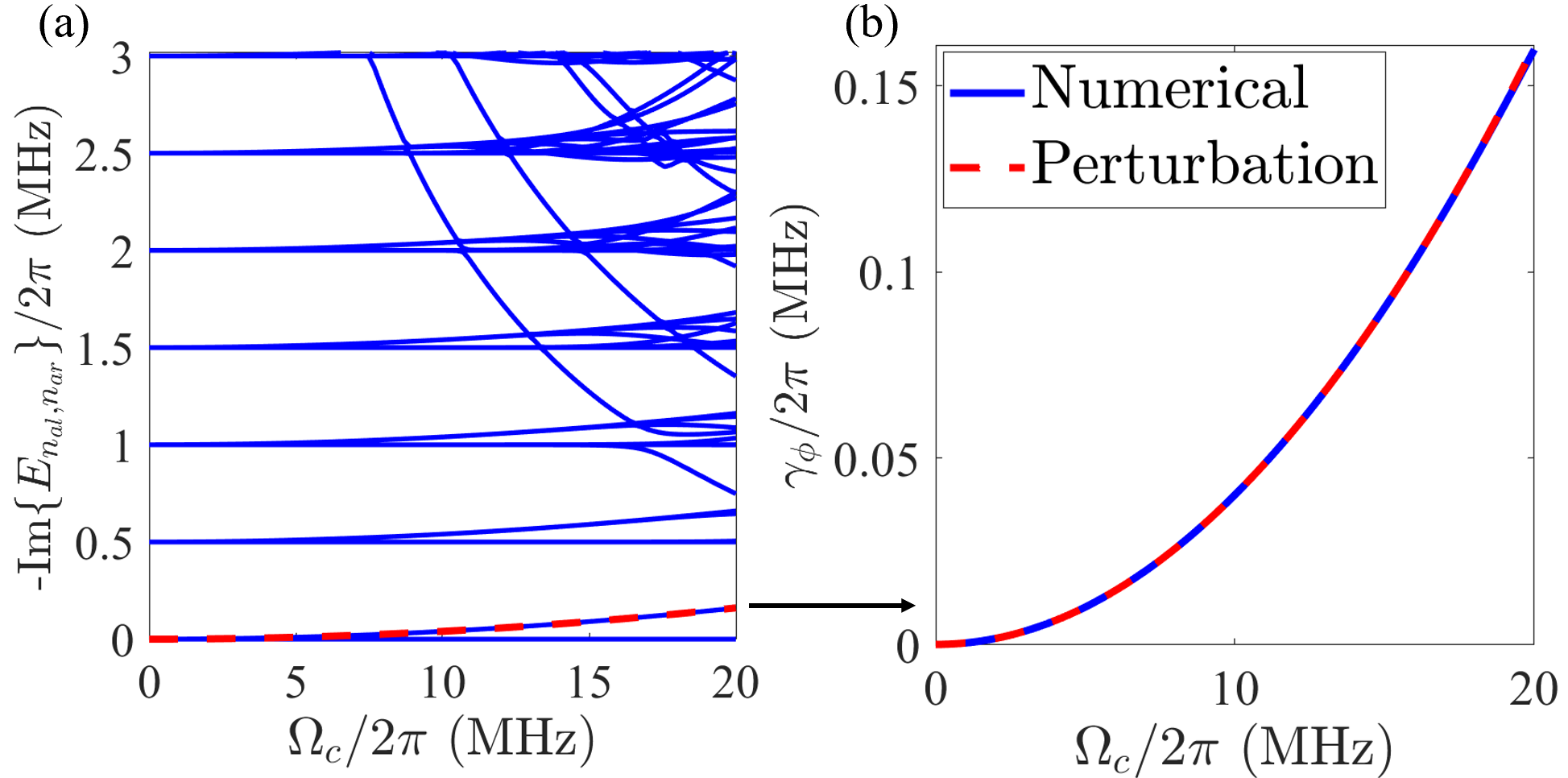}
\caption{(a) Renormalization of the decay rates based on numerical diagonalization. (b) Comparison between numerical and perturbative expression for $\gamma_{\phi}$ in Eq.~(\ref{eqn:MeasIndStSh-gamma_phi Sol}). The simulation is based on numerical diagonalization of $\HO_u$ in Eqs.~(\ref{eqn:EffMap-Def of H_l&H_r})--(\ref{eqn:EffMap-Def of H_kappa}). System parameters are $\Delta_{ad}/2\pi=-2005$, $\Delta_{cd}/2\pi=-5$, $\chi_{ac}/2\pi=-1$ and $\kappa_c/2\pi=1$ MHz. The strongest $\Omega_c$ corresponds to approximately 4 resonator photons. We kept 2 qubit and 14 resonator states in the simulation.}
\label{fig:MeasIndStSh-NumericalDiagOfLindbladian}
\end{figure}

We further validate the SWLPT method with a numerical diagonalization of $\HO_u$ in Eqs.~(\ref{eqn:EffMap-Def of H_l&H_r})--(\ref{eqn:EffMap-Def of H_kappa}). For a constant $\Omega_c$, $\HO_{u}$ is time-independent and can be exactly diagonalized. The real and imaginary parts of the spectrum of $\HO_u$ give the renormalization of the frequencies and dephasing rates for the system. Figure~\ref{fig:MeasIndStSh-NumericalDiagOfLindbladian} shows a numerical sweep of $\Omega_c$ and the corresponding decay rates, where we find that the perturbative expression~(\ref{eqn:MeasIndStSh-Delta_S Sol}) for $\gamma_{\phi}$ captures the low-power dependence very precisely. 

%%%%%%%%%%%%%%%%%%%%% Fig: comparsion with numerics %%%%%%%%%%%%%%%%%%%%%%%%%%%%
%%%%%%%%%%%%%%%%%%%%%%%%%%%%%%%%%%%%%%%%%%%%%%%%%%%%%%%%%%%%%%%%%%%%%%%%%%%%%%%%

%%%%%%%%%%%%%%%%%%%%%%%%%% Sec: Transient behavior %%%%%%%%%%%%%%%%%%%%%%%%%%%%%
\section{Transient behavior of effective interactions}
\label{Sec:Transient}

Our discussion in Sec.~\ref{Sec:MeasIndStSh} was based on a time-independent pulse envelope. The SWLPT method, however, accounts also for the transient dynamics of the effective interactions. Here, we explore evolution under the time-dependent $\HO_{I,\text{eff}}^{\text{ad}}(t)$ of Eq.~(\ref{eqn:EffMap-Adiab H_I,eff}), as well as an adiabatic expansion that brings higher-order corrections in terms of the derivatives of the resonator coherent response $\eta_c(t)$. The latter should be thought of as an independent expansion used in conjunction with SWLPT Eqs.~(\ref{eqn:SWLPT-Def of H_I,eff^(1)})--(\ref{eqn:SWLPT-Def of G_3}). Successive terms in the SWLPT and adiabatic expansions characterize how strong and how fast the interactions (drive) are compared to the transition frequencies. 

%%%%%%%%%%%%%% Fig: Transients %%%%%%%%%%%%%%%%%
\begin{figure}[t!]
\centering
\includegraphics[scale=0.135]{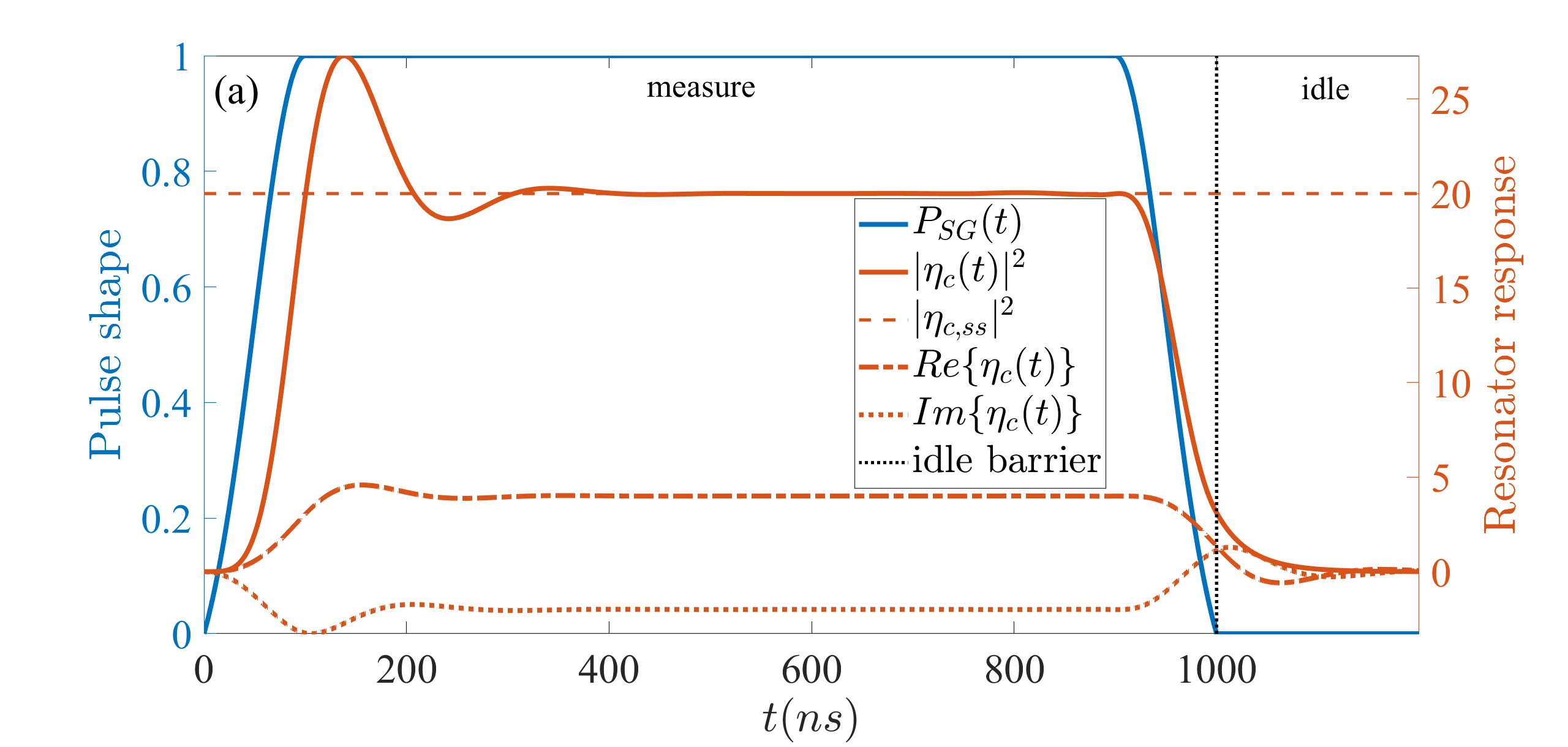}\\
\includegraphics[scale=0.135]{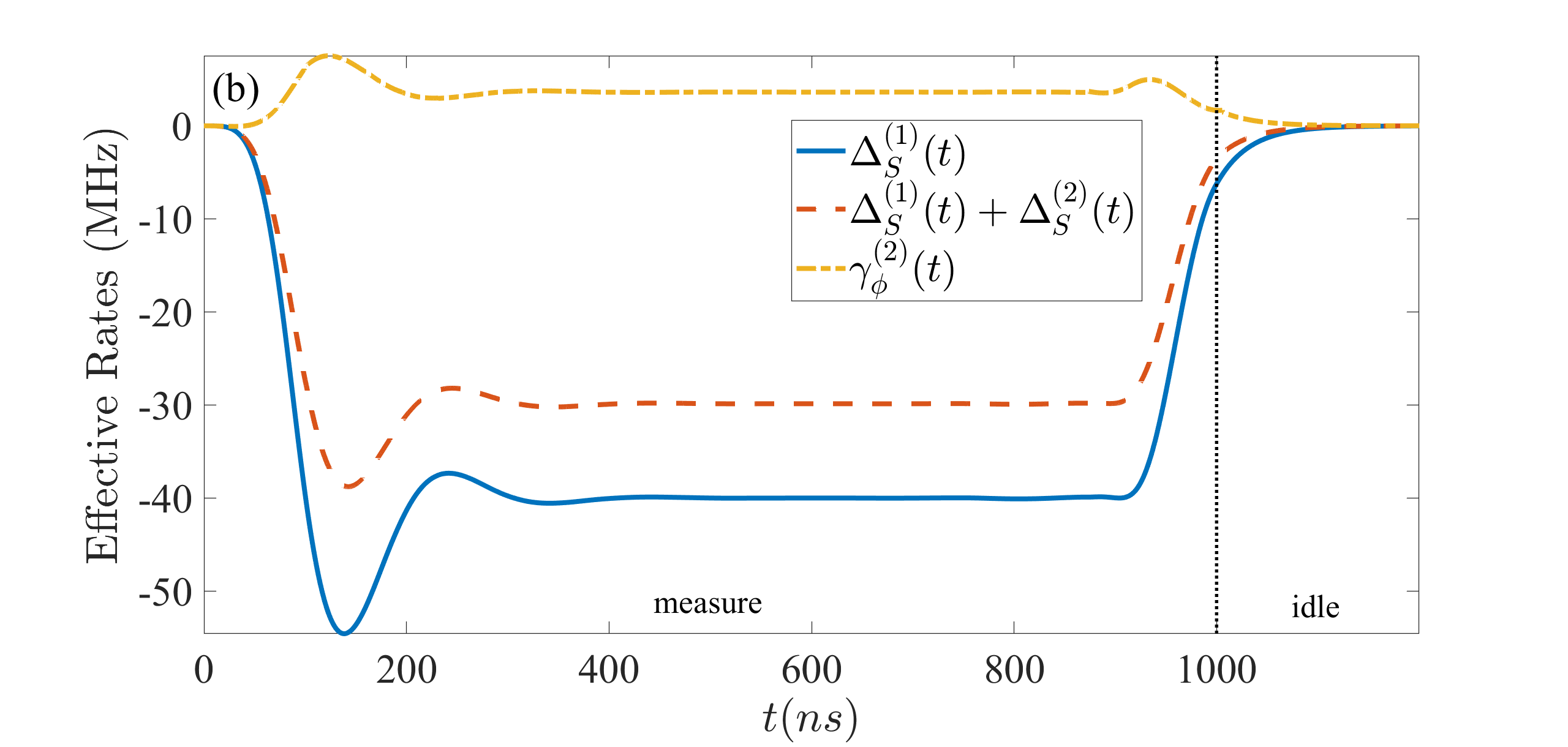}\\
\includegraphics[scale=0.135]{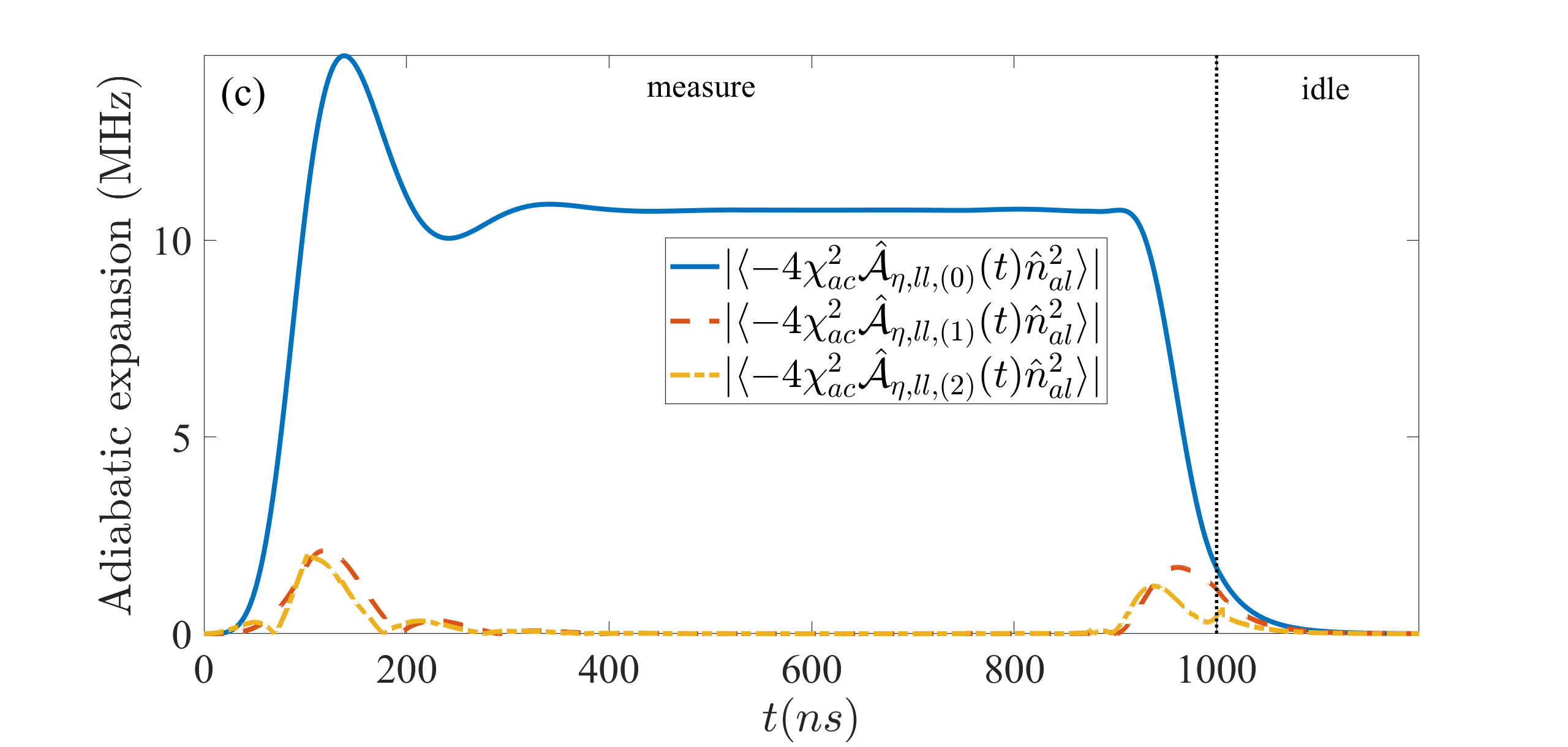}\\
\caption{Transients of the effective interactions based on time-dependent SWLPT. (a) Resonator response and photon number, based on a numerical solution to Eq.~(\ref{eqn:EffMap-Def of eta_c(t)}), for the SG pulse in Eq.~(\ref{eqn:OffRes-Def of SGPulse}). (b) First-order Stark shift, and second-order effective interactions based on the adiabatic expansion in Eq.~(\ref{eqn:Transient-Adiab exp of A_eta,jj}). (c) Comparison of the consequent adiabatic terms, i.e.~the zeroth, first and second derivative in Eq.~(\ref{eqn:Transient-Adiab exp of A_eta,jj}). System and pulse parameters are $\Delta_{ad}/2\pi=-2005$, $\Delta_{cd}/2\pi=-5$, $\chi_{ac}/2\pi=-1$, $\kappa_c/2\pi=5$, $\Omega_c/2\pi=50$ MHz, and $\tau_p=1000$, $\tau_r=100$ and $\sigma_r=50$ ns. The vertical black dotted line marks the end of the SG pulse, beyond which we observe residual resonator occupation. Horizontal dashed line in (a) shows the analytical steady-state photon number that matches the numerical integration.}
\label{fig:Transient-EffTimeDepRates}
\end{figure}
%%%%%%%%%%%%%%%%%%%%%%%%%%%%%%%%%%%%%%%%%%%%%%%%%%%%%%%%%%%%%%%%%%%%%%%%%%%%%

Up to the third order in SWLPT, we find $\HO_{I,\text{eff}}(t)$ in its full time-dependent form as (Appendix~\ref{App:SWLPT})
\begin{align}
\begin{split}
\HO_{I,\text{eff}}(t) &=  2\chi_{ac}|\eta_c(t)|^2 \hat{n}_{al}-2\chi_{ac}|\eta_c(t)|^2 \hat{n}_{ar}\\
&-4\chi_{ac}^2\AO_{\eta,ll}(t)\hat{n}_{al}^2+4\chi_{ac}^2\AO_{\eta,rr}(t)\hat{n}_{ar}^2 \\
&+i4\chi_{ac}^2\kappa_c \bigg[\frac{\hat{\mathcal{B}}_{\eta,lr}(t)}{6}+\frac{\hat{\mathcal{C}}_{\eta,lr}(t)}{2}\bigg]\hat{n}_{al}\hat{n}_{ar} \;,
\end{split}
\label{eqn:Transient-TimeDep H_I,eff(t)}
\end{align}
where $\AO_{\eta,ll}(t)$ and $\AO_{\eta,rr}(t)$ are the left and right second-order time-dependent correlation functions, describing measurement-induced dephasing and further correction to the Stark shift, and $\BO_{\eta,lr}(t)$ and $\CO_{\eta,lr}(t)$ are third-order correlation functions describing the cross (left-right) interaction similar to a collapse term in a Lindblad dissipator. Under adiabatic evolution, we find $\AO_{\eta,ll}^{\text{ad}}(t)=|\eta_c(t)|^2/\hat{\Delta}_{cdl}$, $\AO_{\eta,rr}^{\text{ad}}(t)=|\eta_c(t)|^2/\hat{\Delta}_{cdr}$ and $\BO_{\eta,lr}^{\text{ad}}(t)=\CO_{\eta,lr}^{\text{ad}}(t)=(3/2)|\eta_c(t)|^2/(\hat{\Delta}_{cdl}\hat{\Delta}_{cdr})$, which reduces Eq.~(\ref{eqn:Transient-TimeDep H_I,eff(t)}) to Eq.~(\ref{eqn:EffMap-Adiab H_I,eff}). Full time-dependent solutions for $\BO_{\eta,lr}(t)$ and $\CO_{\eta,lr}(t)$ are involved and we refer the reader to Appendix~\ref{SubApp:SWLPT3}. Here, we discuss the transients of $\AO_{\eta,ll}(t)$. Similar results apply to $\AO_{\eta,rr}(t)$.    

Correlation function $\AO_{\eta,ll}(t)$ describes a second-order effect, generated from a simplified commutator of the form $(i/2)[\GO_1(t),\dot{\GO}_1(t)]$ in Eq.~(\ref{eqn:SWLPT-Def of H_I,eff^(2)}), and is found as (Appendix~\ref{SubApp:SWLPT2})    
\begin{align}
\begin{split}
\hat{\mathcal{A}}_{\eta,ll}(t) & \equiv \frac{1}{2i}\int^{t}dt' \eta_c(t)\eta_c^*(t')e^{i\hat{\Delta}_{cdl}(t-t')} \\
&-\frac{1}{2i}\int^{t}dt' \eta_c^*(t)\eta_c(t')e^{-i\hat{\Delta}_{cdl}(t-t')} \;,
\end{split}
\label{eqn:Transient-Def of A_eta,jj}
\end{align}
which involves the resonator response $\eta_c$ at two different times $t$ and $t'$. Note that the integrals in Eq.~(\ref{eqn:Transient-Def of A_eta,jj}) are indefinite and contract $t'$ into a single time variable in $\AO_{\eta,ll}(t)$. This is understood as $\AO_{\eta,ll}(t)$, and the rest of the correlation functions in Eq.~(\ref{eqn:Transient-TimeDep H_I,eff(t)}), provide the effective \textit{rates}. The effective rotation \textit{angles} are consequently computed by a definite integration of the rates over the pulse duration.      

The adiabatic expression for $\AO_{\eta,ll}(t)$ is the leading-order contribution, found by integrating only over the phase factors in Eq.~(\ref{eqn:Transient-Def of A_eta,jj}). This can be generalized, using integration by parts, resulting in a series in terms of the derivatives of $\eta_c(t)$ and $\eta_c^*(t)$ (Appendix~\ref{SubApp:TransAdiabExp}). Keeping the terms up to $\ddot{\eta}_c(t)$ we find:
\begin{align}
\begin{split}
\AO_{\eta,ll}(t)&=\frac{|\eta_c(t)|^2}{\hat{\Delta}_{cdl}}+\frac{\eta_c(t)\dot{\eta}_c^{*}(t)-\eta_c^*(t)\dot{\eta}_c(t)}{2i \hat{\Delta}_{cdl}^2}\\
&-\frac{\eta_c(t)\ddot{\eta}_c^{*}(t)+\eta_c^*(t)\ddot{\eta}_c(t)}{2 \hat{\Delta}_{cdl}^3}+O\left(\left|\frac{\eta_c(t)\dddot{\eta}_c^*(t)}{\hat{\Delta}_{cdl}^4}\right|\right) \;.
\end{split}
\label{eqn:Transient-Adiab exp of A_eta,jj}
\end{align}
Expansion~(\ref{eqn:Transient-Adiab exp of A_eta,jj}) becomes practical when the pulse ramps are not too sharp compared to the detunings, so that keeping the first few terms is sufficient. An alternative for fast ramps is to use a Fourier representation of the correlation functions (Appendix~\ref{SubApp:TransFourier}).     

We study the transient contributions in Eq.~(\ref{eqn:Transient-Adiab exp of A_eta,jj}) for a Square Gaussian (SG) pulse envelope:  
\begin{align}
P_{\text{SG}}(t)\equiv
\begin{cases}
\frac{e^{-\frac{(t-\tau_r)^2}{2\sigma_r^2}}-e^{-\frac{\tau_r^2}{2\sigma_r^2}}}{1-e^{-\frac{\tau_r^2}{2\sigma_r^2}}}  \;, &0\leq t\leq\tau_r\\
1\;,  &\tau_r \leq t \leq \tau_p-\tau_r\\
\frac{e^{-\frac{[t-(\tau_p-\tau_r)]^2}{2\sigma_r^2}}-e^{-\frac{\tau_r^2}{2\sigma_r^2}}}{1-e^{-\frac{\tau_r^2}{2\sigma_r^2}}} \;, &\tau_p-\tau_r \leq t \leq \tau_p
\end{cases}
\label{eqn:OffRes-Def of SGPulse}
\end{align}
where $\tau_p$, $\tau_r$ and $\sigma_r$ are the pulse time, rise time and the Gaussian standard deviation, respectively. Employing a numerical ODE solver, we obtain $\eta_c(t)$ from Eq.~(\ref{eqn:EffMap-Def of eta_c(t)}) for $\Omega_c(t)=\Omega_c P_{SG}(t)$, as well as its higher-order derivatives. We then substitute the numerical solutions into the analytical transient expressions in Eq.~(\ref{eqn:Transient-Adiab exp of A_eta,jj}).  

%%%%%%%%%%%%%% Fig: Transients %%%%%%%%%%%%%%%%%
\begin{figure}[t!]
\centering
\includegraphics[scale=0.135]{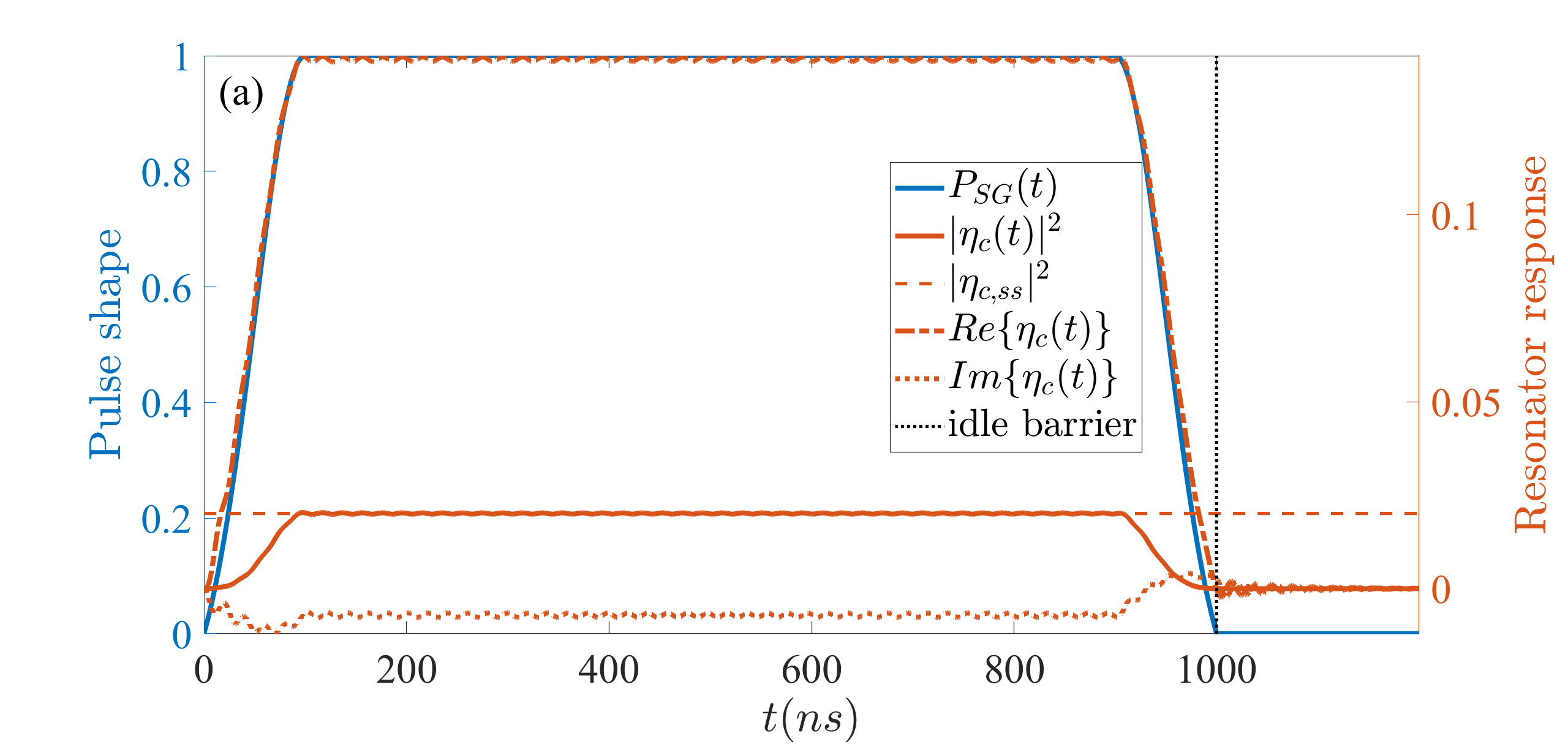}\\
\includegraphics[scale=0.135]{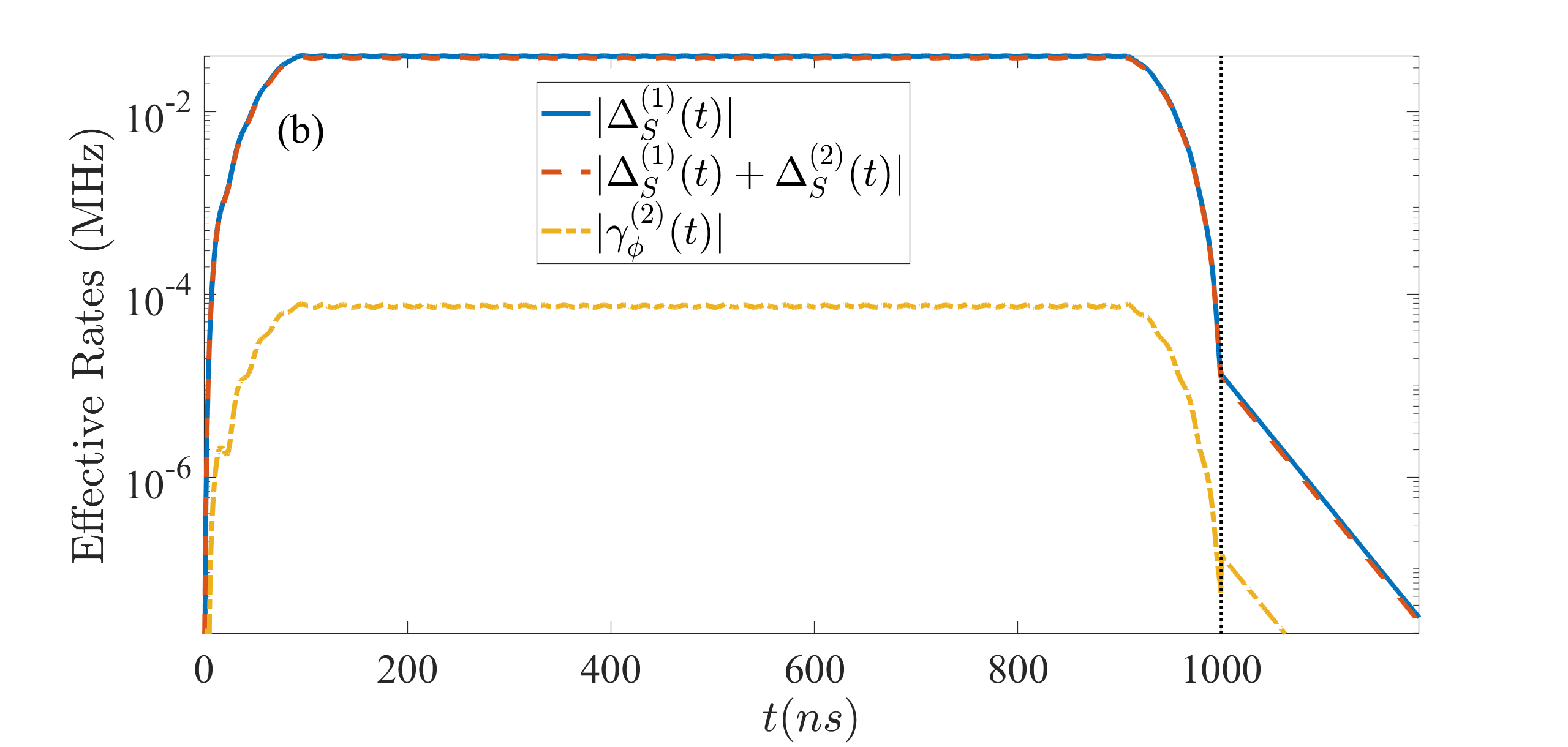}\\
\includegraphics[scale=0.135]{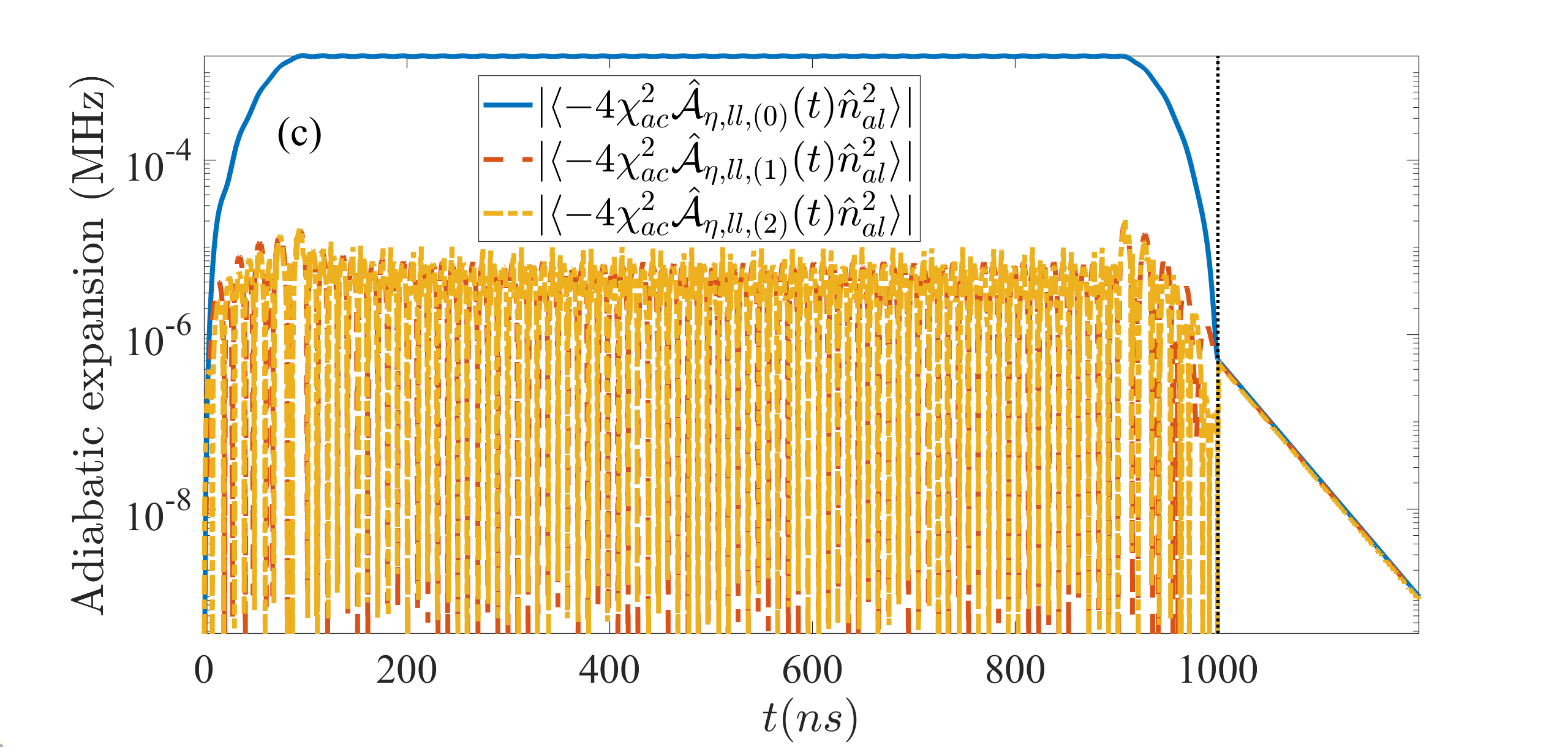}\\
\caption{Transients of the effective interactions for measurement crosstalk. Panels are the same as in Fig.~\ref{fig:Transient-EffTimeDepRates}, except for the log scale in panels (b) and (c). System and pulse parameters are $\Delta_{ad}/2\pi=-2050$, $\Delta_{cd}/2\pi=-50$, $\chi_{ac}/2\pi=-1$, $\kappa_c/2\pi=5$, $\Omega_c/2\pi=14.2$ MHz, and $\tau_p=1000$, $\tau_r=100$ and $\sigma_r=50$ ns. With respect to Fig.~\ref{fig:Transient-EffTimeDepRates}, the detuning and drive amplitude has been changed to -50 MHz, and 14.2 MHz, respectively, equivalent to a photon-transfer cross-talk factor of $10^{-3}$, i.e. 0.02 steady-state photons.}
\label{fig:Transient-EffTimeDepCrossTalk}
\end{figure}
%%%%%%%%%%%%%%%%%%%%%%%%%%%%%%%%%%%%%%%%%%%%%%%%%%%%%%%%%%%%%%%%%%%%%%%%%%%%%

Figure~\ref{fig:Transient-EffTimeDepRates} shows the resonator response, the first- and second-order effective rates, and an adiabatic breakdown of the effective rates in panels (a)--(c), respectively. The system and pulse parameters (see caption) are chosen such that the adiabatic contribution of Eq.~(\ref{eqn:EffMap-Adiab H_I,eff}) is dominant, with weaker corrections from the second and third terms in Eq.~(\ref{eqn:Transient-Adiab exp of A_eta,jj}) that become relevant only during the transient ring up/down of the resonator. The resonator response and the corresponding effective rates demonstrate three stages of (i) ring up with a possible overshoot (dependent on resonator-drive detuning), (ii) relaxation to steady-state, and (iii) ring down with residual occupation that outlasts the control pulse by up to a few hundred nanoseconds.      

We also consider a distinct parameter set in Fig.~\ref{fig:Transient-EffTimeDepCrossTalk}, corresponding to measurement cross-talk, in which photons can leak to a neighboring readout resonator through a shared feed-line, bus, or Purcell filter. A detailed model of such a setup is architecture dependent, and in addition to the two qubits, could include two readout resonators and a resonator mode describing the bus or Purcell filter. Here, however, we assume that the cross-talk is dominated by photon transfer via the bus mode so that we can still model it using the single readout setup, but with modified parameters. Compared to Fig.~\ref{fig:Transient-EffTimeDepRates}, we assume a 50 MHz detuning between the neighboring resonator and drive, and a photon transfer ratio of $0.001$, corresponding to $0.02$ steady-state photons (panel a). In such a scenario, one finds that the second-order effective rates are of the order of 0.1 KHz, so that measurement-induced dephasing on the neighboring qubit is quite weak. However, the first-order Stark shift cross-talk can be tens of KHz (panel b). Moreover, due to the larger detuning, diabatic corrections are suppressed and the effective rates approximately follow the form in Eq.~(\ref{eqn:EffMap-Adiab H_I,eff}) proportional to the instantaneous photon number $|\eta_c(t)|^2$ (panel c).	           
%%%%%%%%%%%%%%%%%%%%%%%%%%%%%%%%%%%%%%%%%%%%%%%%%%%%%%%%%%%%%%%%%%%%%%%%%%%%%%%%

%%%%%%%%%%%%%%%%%%%%%%%%%%% Sec: Conclusion %%%%%%%%%%%%%%%%%%%%%%%%%%%%%%%%%%
\section{Summary and outlook}
\label{Sec:Summary}

We introduced a natural generalization of the SWPT method to be applicable to Lindblad dynamics, a method that we have called SWLPT. Our construction of SWLPT adopts the same effective frame transformations as in the Hamiltonian problem, but applies it to a Schr\"{o}dinger-like (vectorized) equation for the density matrix. The adopted vectorization \cite{Yi_Effective_2001} unfolds the Lindbladian into left and right modes that obey the standard commutation relations, which aligns well with SWLPT that provides corrections in terms of nested commutators. This being said, depending on the problem of interest, other equivalent constructions of SWLPT should be possible, either directly for the denisty matrix, i.e. without vectorization, or using distinct vectorizations.  

To benchmark the SWLPT method, we considered a Kerr-oscillator model for the dispersive measurement of a weakly anharmonic transmon qubit. Applying SWLPT, we derived an effective map that captures the low-power behavior of the Stark shift and measurement-induced dephasing for the qubit, in agreement with earlier literature \cite{Gambetta_Qubit-photon_2006}. The developed SWLPT method is time-dependent, so that it describes the transients of the effective interactions as well. We introduced equivalent representations of the transient correlation functions in the time and Fourier-domains, as well as a representation in the adiabatic limit. In particular, the adiabatic expansion acts as a bridge between time-dependent and time-independent perturbation theories.        

An interesting outcome of employing SWLPT for the transmon readout setup was the possibility of deriving an effective map that is CPTP, and hence a valid quantum channel, under adiabatic response. We note that, unlike SWPT that always leads to an effective unitary time-evolution operator, for SWLPT the CPTP property seems to depend on the choice of the zeroth-order generator and particularly the treatment of the collapse terms. We conjecture that performing a symplectic diagonalization of the collapse term, prior to SWLPT, will lead to CPTP forms for the effective map at arbitrary truncation order. Further study is needed for understanding general properties of the time-dependent SWLPT method beyond our readout example, and to establish under what conditions it or related techniques produce an effective map that is CPTP.    
              
%%%%%%%%%%%%%%%%%%%%%%%%%%%%%%%%%%%%%%%%%%%%%%%%%%%%%%%%%%%%%%%%%%%%%%%%%%%%%

%%%%%%%%%%%%%%%%%%%%%%% Sec: Acknowledgements %%%%%%%%%%%%%%%%%%%%%%%%%%%%%%%
\section{Acknowledgements}
We acknowledge helpful discussions with William Shanks, Ted Thorbeck, Zlatko Minev, David C McKay, Youngseok Kim, David Lokken-Toyli, Isaac Lauer, Oliver Dial, Haggai Landa, Daniel Puzzuoli, and Archana Kamal. Research was sponsored by the Army Research Office and was accomplished under Grant Number W911NF-21-1-0002. The views and conclusions contained in this document are those of the authors and should not be interpreted as representing the official policies, either expressed or implied, of the Army Research Office or the U.S. Government. The U.S. Government is authorized to reproduce and distribute reprints for Government purposes notwithstanding any copyright notation herein.
%%%%%%%%%%%%%%%%%%%%%%%%%%%%%%%%%%%%%%%%%%%%%%%%%%%%%%%%%%%%%%%%%%%%%%%%%%

%%%%%%%%%%%%%%%%%%%%%%%% Appendix A: Vectorization %%%%%%%%%%%%%%%%%%%%%%%
\appendix
\section{Vectorization of Lindblad dynamics}
\label{App:Vector}
We review the correspondence between a given Lindblad equation and an equivalent Schr\"odinger-like equation using a vectorization \cite{Yi_Effective_2001, Prosen_Quantization_2010} in terms of left (original) and right (adjoint) copies of the system Hilbert space. 

Consider a Lindblad evolution for the density matrix $\hat{\rho}(t)$ as 
\begin{align}
\partial_t\hat{\rho}(t)=-i[\hat{H}_s+\hat{H}_d(t),\hat{\rho}(t)]+\sum\limits_{j}\gamma_j \mathcal{D}	[\hat{C}_j]\hat{\rho}(t) \;,
\label{Eq:Vector-Lindblad Eq}
\end{align}
where $\hat{H}_s$ and $\hat{H}_d(t)$ are static and drive Hamiltonian, $\gamma_j$ is the relaxation rate corresponding to the collapse operator $\hat{C}_j$ and $\mathcal{D}[\hat{C}_j]\hat{\rho} \equiv \hat{C}_j\hat{\rho}\hat{C}_j^{\dag}-(1/2)\{\hat{C}_j^{\dag} \hat{C}_j ,\hat{\rho} \}$.

Lindblad Eq.~(\ref{Eq:Vector-Lindblad Eq}) can be mapped into an effective Schr\"odinger-like equation by extending the Hilbert space via an auxiliary copy of the original system that encodes the adjoint states. Given a solution for the density matrix in terms of a system basis $\{\ket{n}\}$ as 
\begin{align}
\hat{\rho}(t)=\sum\limits_{mn} \rho_{mn}(t) \ket{m}\bra{n} \;,
\label{Eq:Vector-Rho expansion}
\end{align}
with $\rho_{mn}(t)\equiv \bra{m} \hat{\rho}(t)\ket{n}$, the corresponding vectorized wavefunction is defined as
\begin{align}
\ket{\Psi_{\hat{\rho}}(t)}=\sum\limits_{mn} \rho_{mn}(t) \ket{m_l}\ket{n_r} \;,  
\label{Eq:Vector-corresponding Psi_rho}
\end{align}
with subscripts $l$ and $r$ denoting the left and the right sectors.

In this vectorization, the Lindblad dynamics in Eq.~(\ref{Eq:Vector-Lindblad Eq}) is equivalent to 
\begin{align}
\partial_t \ket{\Psi_{\hat{\rho}}(t)} = -i \HO_u (t) \ket{\Psi_{\hat{\rho}}(t)}  \;,  
\label{Eq:Vector-Schro-like Eq}
\end{align}
where the extended Hamiltonian $\HO_u(t)$ takes the form:
\begin{align}
& \HO_u(t) \equiv \HO_l(t)-\HO_r(t) + \HO_{\gamma} \;, 
\label{Eq:Vector-Def of Hu}\\
&\HO_{\gamma} \equiv 	\sum\limits_j i \gamma_j \left( \CO_{j,l}\CO_{j,r}-\frac{1}{2}\CO_{j,l}^{\dag} \CO_{j,l} -\frac{1}{2} \CO_{j,r}^{\dag}\CO_{j,r} \right) \;.
\label{Eq:Vector-Def of H_gamma}
\end{align}
Here, $\HO_{l}(t)$ and $\HO_{r}(t)$ represent the left and the right copies of the overall Hamiltonian $\hat{H}_s+\hat{H}_d(t)$, and $\HO_{\gamma}$ is a representation of the dissipators. Note that each dissipator consists of diagonal decay contribution $(-i/2) \gamma_j (\CO_{j,r}^{\dag} \CO_{j,r}+\CO_{j,l}^{\dag} \CO_{j,l})$ and off-diagonal collapse contribution $i \gamma_j \CO_{j,l} \CO_{j,r}$. Therefore, it is only the collapse terms that directly couple the left and right sectors.

In Eqs.~(\ref{Eq:Vector-Def of Hu})--(\ref{Eq:Vector-Def of H_gamma}), the left and right extended operators corresponding to an arbitrary operator $\hat{O}$ are defined according to \cite{Yi_Effective_2001} 
\begin{subequations}
\begin{align}
& \OO_{l}\ket{\Psi_{\hat{\rho}}} \equiv \ket{\Psi_{\hat{O}\hat{\rho}}} \;,
\label{Eq:Vector-Def of MathcalO_l}\\
& \OO_{r}\ket{\Psi_{\hat{\rho}}} \equiv \ket{\Psi_{\hat{\rho}\hat{O}^{\dag}}} \;. 
\label{Eq:Vector-Def of MathcalO_r}
\end{align}
\end{subequations}
Requiring the vectorized notation to give the same matrix elements as the density matrix notation, i.e. $\bra{j}\bra{k}\OO_l \ket{\Psi_{\hat{\rho}}}=\bra{j}\hat{O}\hat{\rho}\ket{k}$ and $\bra{j}\bra{k}\OO_r \ket{\Psi_{\hat{\rho}}}=\bra{j}\hat{\rho}\hat{O}^\dagger\ket{k}$, leads to
\begin{subequations}
\begin{align}
&\OO_l \equiv \hat{O} \otimes \hat{I} \;, 
\label{Eq:Vector-Rep of Ol}\\
&\OO_r \equiv \hat{I} \otimes \hat{O}^*\;.	
\label{Eq:Vector-Rep of Or}
\end{align}
\end{subequations}

Two remarks are in order. First, there is flexibility in the definition of the right extended operator in Eq.~(\ref{Eq:Vector-Def of MathcalO_r}), where it is more common to use $\OO_{r}\ket{\Psi_{\hat{\rho}}} \equiv \ket{\Psi_{\hat{\rho}\hat{O}}}$ \cite{Prosen_Quantization_2010, Mcdonald_Exact_2022}. Our convention leads to the standard commutation relations $[\OO_r,\OO_r^{\dag}]=[\OO_l,\OO_l^{\dag}]$, in contrast to having a minus sign when following the other convention as $[\OO_r,\OO_r^{\dag}]=-[\OO_l,\OO_l^{\dag}]$, and is favorable in terms of bookkeeping given our use of SWLPT which is based on nested commutators (Appendix~\ref{App:SWLPT}).

Second, the vectorization in Eq.~(\ref{Eq:Vector-Schro-like Eq}) follows Ref.~\cite{Yi_Effective_2001} where we work with an equivalent extended Hamiltonian. However, this should only be thought of as a representation of the Linbdladian, i.e. $\LO_u \equiv -i\HO_u$. This choice again is motivated by the use of the same SWPT equations that was developed at the Hamiltonian level \cite{Malekakhlagh_Lifetime_2020, Petrescu_Lifetime_2020, Magesan_Effective_2020, Malekakhlagh_First-Principles_2020, Xiao_Perturbative_2021, Petrescu_Accurate_2021, Malekakhlagh_Mitigating_2022, Malekakhlagh_Optimization_2022}.   

%%%%%%%%%%%%%%%%%%%%%%%%%%%%%%%%%%%%%%%%%%%%%%%%%%%%%%%%%%%%%%%%%%%%%%%%%%%%%%
\section{Displacement transformation of the resonator mode}
\label{App:DispTrans}

In this appendix, we apply a displacement transformation on Eqs.~(\ref{eqn:EffMap-Def of H_l&H_r})--(\ref{eqn:EffMap-Def of H_kappa}) to account for the classical mean-field response of the resonator mode. The displacement transformation is defined as 
\begin{align}
\TO_D[\eta_c(t)] \equiv e^{\eta_c(t)\hat{c}_l^{\dag}-\eta_c^{*}(t)\hat{c}_l}e^{\eta_c^*(t)\hat{c}_r^{\dag}-\eta_c(t)\hat{c}_r} \;,
\label{Eq:DispTrans-Def of D[eta]}
\end{align}
with $\eta_c(t)$ and $\eta_c^*(t)$ as the coherent displacement of the left and the right resonator modes:
\begin{subequations}
\begin{align}
& \TO_{D}^{\dag}[\eta_c(t)] \hat{c}_l \TO_{D}[\eta_c(t)] = \hat{c}_l +\eta_c(t) \;, 
\label{Eq:DispTrans-D^d c_l D}\\
& \TO_{D}^{\dag}[\eta_c(t)] \hat{c}_r \TO_D[\eta_c(t)] = \hat{c}_r +\eta_c^{*}(t) \;.
\label{Eq:DispTrans-D^d c_r D}
\end{align}
\end{subequations}
Moreover, $\TO_{D}[\eta_c(t)]$ is unitary and obeys $\TO_{D}^{-1}[\eta_c(t)]=\TO_{D}^{\dag}[\eta_c(t)]=\TO_{D}[-\eta_c(t)]$.

Using Eqs.~(\ref{Eq:DispTrans-D^d c_l D})--(\ref{Eq:DispTrans-D^d c_r D}), the displaced extended Hamiltonian is obtained as
\begin{align}
\HO_{u,\text{dis}}(t) \equiv \TO_{D}^{\dag}[\eta_c(t)][\HO_u(t)-i\partial_t]\TO_{D}[\eta_c(t)] \;.
\label{Eq:DispTrans-D^d H_u D}
\end{align}
Our first step in the diagonalization of $\HO_u(t)$ is to set the coefficients of the terms that are linear in $\hat{c}_l$, $\hat{c}_l^{\dag}$, $\hat{c}_r$ and $\hat{c}_r^{\dag}$ in Eq.~(\ref{Eq:DispTrans-D^d H_u D}) to zero. The four conditions all result in the same equation for $\eta(t)$ (and similarly for $\eta_c^*(t)$ by complex conjugation) 
\begin{align}
\dot{\eta}_c (t) + \left(i\Delta_{cd}+\frac{\kappa_c}{2}\right)\eta_c(t)=-\frac{i}{2}\Omega_c(t) \;,
\label{Eq:DispTrans-Cond for eta_c(t)}
\end{align}
which is equivalent to the classical response of a driven-dissipative harmonic oscillator under RWA. Based on Eq.~(\ref{Eq:DispTrans-Cond for eta_c(t)}), the steady-state resonator photon number is
\begin{align}
|\eta_{c,\text{ss}}|^2 =\frac{|\Omega_c|^2}{4\left[\Delta_{cd}^2+(\kappa_c/2)^2\right]}\;.
\label{Eq:DispTrans-eta_c,ss Sol}
\end{align}

In the displaced frame, the extended Hamiltonian can be written as
\begin{subequations}
\begin{align}
\begin{split}
\HO_{u,\text{dis}}(t) & \equiv \HO_{l,\text{dis}} (t) - \HO_{r,\text{dis}} (t) +\HO_{\kappa} \;,
\end{split}
\label{Eq:DispTrans-Displaced Hu}
\end{align}
with the left and the right sectors as
\begin{align}
\begin{split}
\HO_{l,\text{dis}}(t) & = \Delta_{ad} \hat{a}_l^{\dag}  \hat{a}_l+\frac{1}{2} \alpha_a \hat{a}_l^{\dag}\hat{a}_l^{\dag}\hat{a}_l\hat{a}_l\\
&+ \Delta_{cd} \hat{c}_l^{\dag}\hat{c}_l+2\chi_{ac}\hat{a}_l^{\dag}\hat{a}_l\hat{c}_l^{\dag}\hat{c}_l\\
& +2\chi_{ac}|\eta_c(t)|^2 \hat{a}_l^{\dag} \hat{a}_l \\
&+ \left[2\chi_{ac} \eta_c^*(t) \hat{a}_l^{\dag}\hat{a}_l\hat{c}_l +\text{H.c.}\right] \;,
\end{split}
\label{Eq:DispTrans-Displaced H_l}\\
\begin{split}
\HO_{r,\text{dis}}(t) & = \Delta_{ad} \hat{a}_r^{\dag}  \hat{a}_r+\frac{1}{2} \alpha_a \hat{a}_r^{\dag}\hat{a}_r^{\dag}\hat{a}_r\hat{a}_r\\
&+ \Delta_{cd} \hat{c}_r^{\dag}\hat{c}_r+2\chi_{ac}\hat{a}_r^{\dag}\hat{a}_r\hat{c}_r^{\dag}\hat{c}_r\\
& +2\chi_{ac}|\eta_c(t)|^2 \hat{a}_r^{\dag} \hat{a}_r \\
&+ \left[2\chi_{ac} \eta_c(t)\hat{a}_r^{\dag}\hat{a}_r\hat{c}_r +\text{H.c.}\right] \;.
\end{split}
\label{Eq:DispTrans-Displaced H_r}
\end{align}
\end{subequations}
Note that the displaced Hamiltonian~(\ref{Eq:DispTrans-Displaced Hu}) is diagonal with respect to the qubit subspace, but off-diagonal due to two distinct contributions. The first is a time-independent dissipative coupling due to the collapse term $i \kappa_c \hat{c}_r\hat{c}_l$, and the second is the time-dependent nonlinear number-quadrature coupling terms in the last line of Eqs.~(\ref{Eq:DispTrans-Displaced H_l})--(\ref{Eq:DispTrans-Displaced H_r}). The goal of SWLPT is to derive an effective (block) diagonal model that accounts for the effects of these off-diagonal terms perturbatively.
%%%%%%%%%%%%%%%%%%%%%%%%%%%%%%%%%%%%%%%%%%%%%%%%%%%%%%%%%%%%%%%%%%%%%%%%%%%%%%%%

\section{Schrieffer-Wolff-Lindblad Perturbation Theory for dispersive readout}
\label{App:SWLPT}
In this appendix, we use time-dependent SWLPT to diagonalize the displaced extended Hamiltonian~(\ref{Eq:DispTrans-Displaced Hu}). We treat both off-diagonal interaction terms on equal footing by keeping them in the extended interaction Hamiltonian
\begin{align}
\begin{split}
\HO_{\text{int}}(t) &= 2\chi_{ac}|\eta_c(t)|^2 \hat{a}_l^{\dag} \hat{a}_l-2\chi_{ac}|\eta_c(t)|^2 \hat{a}_r^{\dag} \hat{a}_r\\
&+\Big[2\chi_{ac} \eta_c^*(t) \hat{a}_l^{\dag}\hat{a}_l \hat{c}_l
+2\chi_{ac} \eta_c(t) \hat{a}_l^{\dag}\hat{a}_l \hat{c}_l^{\dag}\Big]\\
&-\Big[2\chi_{ac} \eta_c(t)\hat{a}_r^{\dag}\hat{a}_r \hat{c}_r+2\chi_{ac} \eta_c^*(t)\hat{a}_r^{\dag}\hat{a}_r \hat{c}_r^{\dag}\Big]\\
&+i\kappa_c \hat{c}_l\hat{c}_r \;.
\end{split}
\label{Eq:SWLPT-Def of Hint(t)}
\end{align}
Time-independent diagonal terms are accounted for in the zeroth-order extended Hamiltonian $\HO_0 \equiv \HO_{l0} - \HO_{r0}$:
\begin{subequations}
\begin{align}
\begin{split}
\HO_{l0} & \equiv \Delta_{ad} \hat{a}_l^{\dag}  \hat{a}_l+\frac{1}{2} \alpha_a \hat{a}_l^{\dag}\hat{a}_l^{\dag}\hat{a}_l\hat{a}_l\\
&+ \left(\Delta_{cd}-i\frac{\kappa_c}{2}\right) \hat{c}_l^{\dag}\hat{c}_l+2\chi_{ac}\hat{a}_l^{\dag}\hat{a}_l\hat{c}_l^{\dag}\hat{c}_l \;,
\end{split}
\label{Eq:SWLPT-Def of H_l0}\\
\begin{split}
\HO_{r0} & \equiv \Delta_{ad} \hat{a}_r^{\dag}  \hat{a}_r+\frac{1}{2} \alpha_a \hat{a}_r^{\dag}\hat{a}_r^{\dag}\hat{a}_r\hat{a}_r\\
&+ \left(\Delta_{cd}+i\frac{\kappa_c}{2}\right) \hat{c}_r^{\dag}\hat{c}_r+2\chi_{ac}\hat{a}_r^{\dag}\hat{a}_r\hat{c}_r^{\dag}\hat{c}_r \;.
\end{split}
\label{Eq:SWLPT-Def of H_r0}
\end{align}
\end{subequations}
Note that a different grouping of terms is possible. Keeping the collapse term $i\kappa_c \hat{c}_l \hat{c}_r$ in $\HO_0$ ensures the Lindblad form for both $\HO_0$ and $\HO_{\text{int}}(t)$. The tradeoff, however, is an off-diagonal zeroth-order generator $\HO_0$, that can be diagonalized using symplectic transformations \cite{Prosen_Quantization_2010, Mcdonald_Exact_2022} (see the end of Sec.~\ref{Sec:SWLPT} for further discussion).

To simplify the perturbation, we work in the interaction frame with respect to $\HO_0$ via the similarity transformation:
\begin{align}
\begin{split}
\HO_I(t) &\equiv e^{i \HO_0 t} [\HO_{0}+\HO_{\text{int}}(t)-i\partial_t] e^{-i \HO_0 t}\\
&=e^{i \HO_0 t} \HO_{\text{int}}(t) e^{-i \HO_0 t} \;.
\label{Eq:SWLPT-Def of HI(t)}
\end{split}
\end{align}
Note that $\HO_0$ is \textit{not} Hermitian due to the diagonal terms in the dissipator. Therefore, $\exp(\pm i \HO_0 t)$ is not a unitary transformation. More specifically, $\exp(+i \HO_0 t)$ and $\exp(-i \HO_0 t)$ contain terms that gain (decay) in time.\\

Since $\HO_{\text{int}}(t)$ is off-diagonal \textit{only} with respect to the resonator operators, it is sufficient to obtain the following interaction-frame transformations: 
\begin{subequations}
\begin{align}
& e^{i \HO_0 t} \hat{c}_l e^{-i \HO_0 t} = e^{-i \hat{\Delta}_{cdl}t}\hat{c}_l \;,
\label{Eq:SWLPT-IntForm of cL}\\
& e^{i \HO_0 t} \hat{c}_l^{\dag} e^{-i \HO_0 t} = \hat{c}_l^{\dag} e^{i \hat{\Delta}_{cdl}t} \;,  
\label{Eq:SWLPT-IntForm of cL^d}\\
& e^{i \HO_0 t} \hat{c}_r e^{-i \HO_0 t} = e^{i \hat{\Delta}_{cdr}t}\hat{c}_r \;,
\label{Eq:SWLPT-IntForm of cR}\\
& e^{i \HO_0 t} \hat{c}_r^{\dag} e^{-i \HO_0 t} = \hat{c}_r^{\dag} e^{-i \hat{\Delta}_{cdr}t}  \;,
\label{Eq:SWLPT-IntForm of cR^d}
\end{align}
\end{subequations}
with the operator-valued detunings $\hat{\Delta}_{cdl}$ and $\hat{\Delta}_{cdr}$ defined as
\begin{subequations}
\begin{align}
&\hat{\Delta}_{cdl} \equiv \Delta_{cd}-i\frac{\kappa_c}{2}+2\chi_{ac}\hat{a}_l^{\dag}\hat{a}_l \;,
\label{Eq:SWLPT-Def of Delta_cdl}\\
&\hat{\Delta}_{cdr} \equiv \Delta_{cd}+i\frac{\kappa_c}{2}+2\chi_{ac}\hat{a}_r^{\dag}\hat{a}_r \;.
\label{Eq:SWLPT-Def of Delta_cdr}
\end{align}
\end{subequations}
Using Eqs.~(\ref{Eq:SWLPT-IntForm of cL})--(\ref{Eq:SWLPT-Def of Delta_cdr}), the interaction-frame extended Hamiltonian $\HO_I(t)$ is found as
\begin{align}
\begin{split}
&\HO_{I}(t) = 2\chi_{ac}|\eta_c(t)|^2 \hat{a}_l^{\dag} \hat{a}_l-2\chi_{ac}|\eta_c(t)|^2 \hat{a}_r^{\dag} \hat{a}_r\\
&+\Big[2\chi_{ac} \eta_c^*(t) \hat{a}_l^{\dag}\hat{a}_l \hat{c}_l e^{-i \hat{\Delta}_{cdl}t}+2\chi_{ac} \eta_c(t) \hat{a}_l^{\dag}\hat{a}_l\hat{c}_l^{\dag}e^{i \hat{\Delta}_{cdl}t}\Big]\\
&-\Big[2\chi_{ac} \eta_c(t)\hat{a}_r^{\dag}\hat{a}_r \hat{c}_r e^{i \hat{\Delta}_{cdr}t} +2\chi_{ac} \eta_c^*(t)\hat{a}_r^{\dag}\hat{a}_r \hat{c}_r^{\dag}e^{-i \hat{\Delta}_{cdr}t}\Big]\\
&+ i\kappa_c \hat{c}_l \hat{c}_r e^{-i\hat{\Delta}_{cdl}t}e^{+i\hat{\Delta}_{cdr}t} \;.
\end{split}
\label{Eq:SWLPT-H_I(t)}
\end{align}

\subsection{First order}
\label{SubApp:SWLPT1}

Based on Eq.~(\ref{eqn:SWLPT-Def of H_I,eff^(1)}), up to the first order, the effective extended Hamiltonian contains the $2\chi_{ac}$-per-photon Stark shift for the left and the right qubit modes as
\begin{align}
\HO_{I,\text{eff}}^{(1)}(t)= 2\chi_{ac}|\eta_c(t)|^2 \hat{a}_l^{\dag} \hat{a}_l-2\chi_{ac}|\eta_c(t)|^2 \hat{a}_r^{\dag} \hat{a}_r \;.
\label{Eq:SWLPT-H_I,eff^(1) Sol}
\end{align}
According to Eq.~(\ref{eqn:SWLPT-Def of G_1}), the first-order generator $\GO_1(t)$ is found as the indefinite integral of the rest of the off-diagonal terms in $\HO_I(t)$ as  
\begin{align}
\begin{split}
\GO_1(t) & =  \int^{t}dt' 2\chi_{ac} \eta_c^*(t') \hat{a}_l^{\dag}\hat{a}_l \hat{c}_l e^{-i \hat{\Delta}_{cdl}t'} \\
&+\int^{t}dt' 2\chi_{ac} \eta_c(t') \hat{a}_l^{\dag}\hat{a}_l \hat{c}_l^{\dag}e^{i \hat{\Delta}_{cdl}t'}\\
&-\int^{t}dt'2\chi_{ac} \eta_c(t')\hat{a}_r^{\dag}\hat{a}_r \hat{c}_r e^{i \hat{\Delta}_{cdr}t'} \\
&-\int^{t}dt' 2\chi_{ac} \eta_c^*(t')\hat{a}_r^{\dag}\hat{a}_r \hat{c}_r^{\dag}e^{-i \hat{\Delta}_{cdr}t'}\\
&+\int^{t}dt' i\kappa_c \hat{c}_l \hat{c}_r e^{-i\hat{\Delta}_{cdl}t'}e^{+i\hat{\Delta}_{cdr}t'} \;.
\end{split}
\label{Eq:SWLPT-G1 Sol}
\end{align}

\subsection{Second order}
\label{SubApp:SWLPT2}
Note that since $\HO_{I,\text{eff}}^{(1)}(t)$ is independent of the resonator mode, and $\hat{\mathcal{G}}_1(t)$ is off-diagonal only with respect to the resonator modes, we find $[\hat{\mathcal{G}}_1(t),\HO_{I,\text{eff}}^{(1)}(t)]=0$, from which we conclude that $[\hat{\mathcal{G}}_1(t),\HO_{I}(t)]=[\hat{\mathcal{G}}_1(t),\dot{\hat{\mathcal{G}}}_1(t)]$. This simplifies the second-order SWLPT equations~(\ref{eqn:SWLPT-Def of H_I,eff^(2)})--(\ref{eqn:SWLPT-Def of G_2}) to 
\begin{subequations}
\begin{align}
&\HO_{I,\text{eff}}^{(2)}(t)=\mathcal{S} \Big(\frac{i}{2} [\GO_1(t),\dot{\GO}_1(t)] \Big) \;,
\label{Eq:SWLPT-SimpleDef of H_I,eff^(2)}\\
&\dot{\GO}_{2}(t)=\mathcal{N} \Big(\frac{i}{2} [\GO_1(t),\dot{\GO}_1(t)]\Big) \;.
\label{Eq:SWLPT-SimpleDef of G_2}
\end{align}
\end{subequations}

Substituting Eq.~(\ref{Eq:SWLPT-G1 Sol}) into Eq.~(\ref{Eq:SWLPT-SimpleDef of H_I,eff^(2)}) and further simplifying gives
\begin{subequations}
\begin{align}
\begin{split}
\HO_{I,\text{eff}}^{(2)}(t) =&-4\chi_{ac}^2\hat{\mathcal{A}}_{\eta,ll}(t)(\hat{a}_l^{\dag}\hat{a}_l)^2\\
&+4\chi_{ac}^2\hat{\mathcal{A}}_{\eta,rr}(t)(\hat{a}_r^{\dag}\hat{a}_r)^2 \;,
\end{split}
\label{Eq:SWLPT-H_I,eff^(2) Sol}
\end{align}
where we have organized the contributions in terms of the left-only and the right-only sectors. The correponding second-order operator-valued correlation functions $\hat{\mathcal{A}}_{\eta,ll}(t)$ and $\hat{\mathcal{A}}_{\eta,rr}(t)$ encode the dependence of the effective interactions on the drive through $\eta_c(t)$ as
\begin{align}
\begin{split}
\hat{\mathcal{A}}_{\eta,ll}(t) & \equiv  \frac{1}{2i}\int^{t}dt' \eta_c(t)\eta_c^*(t')e^{i\hat{\Delta}_{cdl}(t-t')} \\
&-\frac{1}{2i}\int^{t}dt' \eta_c^*(t)\eta_c(t')e^{-i\hat{\Delta}_{cdl}(t-t')}\;,
\end{split}
\label{Eq:SWLPT-Def of A_eta,ll}
\end{align}
\begin{align}
\begin{split}
\hat{\mathcal{A}}_{\eta,rr}(t) & \equiv \frac{1}{2i}\int^{t}dt' \eta_c(t)\eta_c^*(t')e^{i\hat{\Delta}_{cdr}(t-t')} \\
&-\frac{1}{2i}\int^{t}dt' \eta_c^*(t)\eta_c(t')e^{-i\hat{\Delta}_{cdr}(t-t')} \;.
\end{split}
\label{Eq:SWLPT-Def of A_eta,rr}
\end{align}
\end{subequations}

%%%%%%%%%%%%%%%%%%%%%%%%%%%%%%%%%%%%%%%%%%%%%%%%%%%%%%%%%%%%%%%%%%%%%%%

Moreover, we find $\hat{\mathcal{G}}_2(t)$ as the off-diagonal contributions according to Eq.~(\ref{Eq:SWLPT-SimpleDef of G_2}) as
\begin{align}
\begin{split}
&\hat{\mathcal{G}}_2(t) = \\
&-\frac{1}{2}\int^{t}dt'\int^{t'}dt'' 2\chi_{ac}\kappa_c \eta_c(t')  \hat{a}_l^{\dag}\hat{a}_l \hat{c}_r e^{i \hat{\Delta}_{cdl}(t'-t'')} e^{i\hat{\Delta}_{cdr}t''}\\
&+\frac{1}{2}\int^{t}dt'\int^{t'}dt'' 2\chi_{ac}\kappa_c \eta_c(t'') \hat{a}_l^{\dag}\hat{a}_l \hat{c}_r e^{-i \hat{\Delta}_{cdl}(t'-t'')} e^{i\hat{\Delta}_{cdr}t'} \\
&+\frac{1}{2}\int^{t}dt'\int^{t'}dt'' 2\chi_{ac}\kappa_c \eta_c^*(t') \hat{a}_r^{\dag}\hat{a}_r \hat{c}_l e^{-i \hat{\Delta}_{cdr}(t'-t'')} e^{-i\hat{\Delta}_{cdl}t''}\\
&-\frac{1}{2}\int^{t}dt'\int^{t'}dt'' 2\chi_{ac}\kappa_c \eta_c^*(t'') \hat{a}_r^{\dag}\hat{a}_r \hat{c}_l e^{i \hat{\Delta}_{cdr}(t'-t'')} e^{-i\hat{\Delta}_{cdl}t'}.
\end{split}
\label{Eq:SWLPT-G2 Sol}
\end{align}

Note that, up to the second order, the collapse term [last line of Eq.~(\ref{Eq:SWLPT-G1 Sol})] does not lead to a renormalization of the effective Hamiltonian. It will, however, appear at the third-order SWLPT through nested commutators.  

\subsection{Third order}
\label{SubApp:SWLPT3}
Using similar relations as for the second order, we first simplify the third-order effective SWLPT equations. Given the form of $\HO_I(t)$, $\GO_1(t)$, and $\GO_2(t)$ in Eqs.~(\ref{Eq:SWLPT-H_I(t)}), (\ref{Eq:SWLPT-G1 Sol}), and (\ref{Eq:SWLPT-G2 Sol}), one finds $[\hat{\mathcal{G}}_1(t),\HO_{I}(t)]=[\hat{\mathcal{G}}_1(t),\dot{\hat{\mathcal{G}}}_1(t)]$, $[\hat{\mathcal{G}}_2(t),\HO_{I}(t)]=[\hat{\mathcal{G}}_2(t),\dot{\hat{\mathcal{G}}}_1(t)]$, and $(i/2)[\hat{\mathcal{G}}_1(t),[\hat{\mathcal{G}}_1(t),\dot{\hat{\mathcal{G}}}_1(t)]]=[\hat{\mathcal{G}}_1(t),\dot{\hat{\mathcal{G}}}_2(t)]$. Using these relations Eq.~(\ref{eqn:SWLPT-Def of H_I,eff^(3)}) reduces to 
\begin{align}
\begin{split}
\hat{\mathcal{H}}_{\text{I,eff}}^{(3)}(t)&=\mathcal{S}\Big(\frac{i}{6}[\hat{\mathcal{G}}_1(t),\dot{\hat{\mathcal{G}}}_2(t)]+\frac{i}{2}[\hat{\mathcal{G}}_2(t),\dot{\hat{\mathcal{G}}}_1(t)]\Big)\;.
\end{split}
\label{Eq:SWLPT-SimpleDef of H_I,eff^(3)}
\end{align}

Both contributions in Eq.~(\ref{Eq:SWLPT-SimpleDef of H_I,eff^(3)}) lead to a cross interaction between the left and the right sectors proportional to $(\hat{a}_l^{\dag}\hat{a}_l)(\hat{a}_r^{\dag}\hat{a}_r)$. In particular, the first term can be written compactly as
\begin{align}
\begin{split}
\mathcal{S}\Big(\frac{i}{6}[\hat{\mathcal{G}}_1(t),\dot{\hat{\mathcal{G}}}_2(t)]\Big)= \frac{i}{6}(2\chi_{ac})^2\kappa_c \hat{\mathcal{B}}_{\eta,lr}(t)(\hat{a}_l^{\dag}\hat{a}_l)(\hat{a}_r^{\dag}\hat{a}_r) \;,
\end{split}
\label{Eq:SWLPT-1st Term of H_I,eff^(3)}
\end{align}
with the third-order correlation function $\hat{\mathcal{B}}_{\eta,lr}(t)$ defined as
\begin{align}
\begin{split}
&\hat{\mathcal{B}}_{\eta,lr}(t)\equiv \\
&-\frac{1}{2}\int^{t}dt'\int^{t}dt'' \eta_c^* (t') \eta_c(t) e^{i \hat{\Delta}_{cdl}(t-t'')} e^{-i \hat{\Delta}_{cdr}(t'-t'')} \\
&+\frac{1}{2}\int^{t}dt'\int^{t}dt'' \eta_c^* (t') \eta_c(t'') e^{-i \hat{\Delta}_{cdl}(t-t'')} e^{i \hat{\Delta}_{cdr}(t-t')}\\
&-\frac{1}{2}\int^{t}dt'\int^{t}dt'' \eta_c(t') \eta_c^*(t) e^{i \hat{\Delta}_{cdl}(t'-t'')} e^{-i \hat{\Delta}_{cdr}(t-t'')}\\
&+\frac{1}{2}\int^{t}dt'\int^{t}dt'' \eta_c(t') \eta_c^*(t'') e^{-i \hat{\Delta}_{cdl}(t-t')} e^{i \hat{\Delta}_{cdr}(t-t'')} \;.
\end{split}
\label{Eq:SWLPT-Def of B_eta,lr}
\end{align}
The second term in Eq.~(\ref{Eq:SWLPT-SimpleDef of H_I,eff^(3)}) takes a similar form
\begin{align}
\begin{split}
\mathcal{S}\Big(\frac{i}{2}[\hat{\mathcal{G}}_2(t),\dot{\hat{\mathcal{G}}}_1(t)]\Big)= \frac{i}{2}(2\chi_{ac})^2\kappa_c \hat{\mathcal{C}}_{\eta,lr}(t)(\hat{a}_l^{\dag}\hat{a}_l)(\hat{a}_r^{\dag}\hat{a}_r) \;,
\end{split}
\label{Eq:SWLPT-2nd Term of H_I,eff^(3)}
\end{align}
with a distinct third-order correlation function $\hat{\mathcal{C}}_{\eta,lr}(t)$ as
\begin{align}
\begin{split}
&\hat{\mathcal{C}}_{\eta,lr}(t)\equiv \\
&+\frac{1}{2}\int^{t}dt'\int^{t'}dt'' \eta_c(t') \eta_c^*(t) e^{i \hat{\Delta}_{cdl}(t'-t'')} e^{-i\hat{\Delta}_{cdr}(t-t'')}\\
&-\frac{1}{2}\int^{t}dt'\int^{t'}dt'' \eta_c(t'') \eta_c^*(t) e^{-i \hat{\Delta}_{cdl}(t'-t'')} e^{-i\hat{\Delta}_{cdr}(t-t')} \\
&+\frac{1}{2}\int^{t}dt'\int^{t'}dt'' \eta_c^*(t') \eta_c(t) e^{i\hat{\Delta}_{cdl}(t-t'')} e^{-i \hat{\Delta}_{cdr}(t'-t'')}\\
&-\frac{1}{2}\int^{t}dt'\int^{t'}dt'' \eta_c^*(t'') \eta_c(t) e^{i\hat{\Delta}_{cdl}(t-t')} e^{i \hat{\Delta}_{cdr}(t'-t'')}  \;.
\end{split}
\label{Eq:SWLPT-Def of C_eta,lr}
\end{align}
Putting the contributions together, the third-order effective Hamiltonian reads
\begin{align}
\begin{split}
\HO_{I,\text{eff}}^{(3)}(t) &= \frac{i}{6}(2\chi_{ac})^2\kappa_c \hat{\mathcal{B}}_{\eta,lr}(t)(\hat{a}_l^{\dag}\hat{a}_l)(\hat{a}_r^{\dag}\hat{a}_r) \\
&+\frac{i}{2}(2\chi_{ac})^2\kappa_c \hat{\mathcal{C}}_{\eta,lr}(t)(\hat{a}_l^{\dag}\hat{a}_l)(\hat{a}_r^{\dag}\hat{a}_r) \;.
\end{split}
\label{Eq:SWLPT-H_I,eff^(3) Sol}
\end{align}

\subsection{Adiabatic approximation}
\label{SubApp:SWLPTAdiabExp}
We next discuss adiabaticity and provide adiabatic approximations to the effective interactions in Eqs.~(\ref{Eq:SWLPT-Def of A_eta,ll}), (\ref{Eq:SWLPT-Def of A_eta,rr}), (\ref{Eq:SWLPT-Def of B_eta,lr}) and~(\ref{Eq:SWLPT-Def of C_eta,lr}). For instance, consider the second-order correlation $\hat{\mathcal{A}}_{\eta,ll}(t)$ in Eq.~(\ref{Eq:SWLPT-Def of A_eta,ll}). We can apply an adiabatic expansion in $\dot{\eta}_c^*(t)/\hat{\Delta}_{cdl}$ via integration by parts: 
\begin{align}
\begin{split}
\hat{\mathcal{A}}_{\eta,ll} &= \frac{1}{2i}\Big[\frac{|\eta_c(t)|^2}{-i\hat{\Delta}_{cdl}}-\frac{|\eta_c(t)|^2}{i\hat{\Delta}_{cdl}}\Big]+ O\left(\frac{\eta_c(t)\dot{\eta}^*_c(t)}{\hat{\Delta}_{cdl}^2}\right) \\
&\equiv \frac{|\eta_c(t)|^2}{\hat{\Delta}_{cdl}}+ O\left(\frac{\eta_c(t)\dot{\eta}^*_c(t)}{\hat{\Delta}_{cdl}^2}\right) \;.
\end{split}
\label{Eq:SWLPT-Adiab App of IM}
\end{align}
Therefore, under adiabatic evolution, $\hat{\mathcal{A}}_{\eta,ll}(t)$ takes the form of photon number over $\hat{\Delta}_{cdl}$ and is valid when $|\dot{\eta}_c(t)|\ll |\braket{\hat{\Delta}_{cdl}}|$.

Applying adiabatic approximation to the correlation functions in Eqs.~(\ref{Eq:SWLPT-Def of A_eta,ll})--(\ref{Eq:SWLPT-Def of A_eta,rr}),~(\ref{Eq:SWLPT-Def of B_eta,lr}) and~(\ref{Eq:SWLPT-Def of C_eta,lr}) we find
\begin{subequations}
\begin{align}
&\hat{\mathcal{A}}_{\eta,ll}^{\text{ad}}(t) = \frac{|\eta_c(t)|^2}{\hat{\Delta}_{cdl}}\;,
\label{Eq:SWLPT-Adiab A_eta,ll}\\
&\hat{\mathcal{A}}_{\eta,rr}^{\text{ad}}(t) = \frac{|\eta_c(t)|^2}{\hat{\Delta}_{cdr}}\;,
\label{Eq:SWLPT-Adiab A_eta,rr}\\
&\hat{\mathcal{B}}_{\eta,lr}^{\text{ad}}(t) = \frac{3|\eta_c(t)|^2}{2\hat{\Delta}_{cdl}\hat{\Delta}_{cdr}}\;,
\label{Eq:SWLPT-Adiab B_eta,rr}\\
&\hat{\mathcal{C}}_{\eta,lr}^{\text{ad}}(t) = \frac{3|\eta_c(t)|^2}{2\hat{\Delta}_{cdl}\hat{\Delta}_{cdr}}\;,
\label{Eq:SWLPT-Adiab C_eta,rr}
\end{align}
\end{subequations}
where the superscript ``$\text{ad}$'' denotes the lowest order adiabatic term. Employing the adiabatic expressions~(\ref{Eq:SWLPT-Adiab A_eta,ll})--(\ref{Eq:SWLPT-Adiab C_eta,rr}), $\HO_{I,\text{eff}}^{\text{ad}}(t)$ can be written compactly as
\begin{align}
\begin{split}
\HO_{I,\text{eff}}^{\text{ad}}(t)  = &+ 2\chi_{ac}|\eta_c(t)|^2 \hat{a}_l^{\dag} \hat{a}_l-2\chi_{ac}|\eta_c(t)|^2 \hat{a}_r^{\dag} \hat{a}_r \\
&-\frac{4\chi_{ac}^2|\eta_c(t)|^2}{\hat{\Delta}_{cdl}} (\hat{a}_l^{\dag}\hat{a}_l)^2 +\frac{4\chi_{ac}^2|\eta_c(t)|^2}{\hat{\Delta}_{cdr}} (\hat{a}_r^{\dag}\hat{a}_r)^2 \\ 
&+i \frac{4\chi_{ac}^2 \kappa_c |\eta_c(t)|^2}{\hat{\Delta}_{cdl} \hat{\Delta}_{cdr}} (\hat{a}_l^{\dag}\hat{a}_l) (\hat{a}_r^{\dag}\hat{a}_r) \;.\\ 
\end{split}
\label{Eq:SWLPT-Adiab H_I,eff}
\end{align}
Alternatively, inserting the explicit expressions for $\hat{\Delta}_{cdl}$ and $\hat{\Delta}_{cdr}$ in Eqs.~(\ref{Eq:SWLPT-Def of Delta_cdl})--(\ref{Eq:SWLPT-Def of Delta_cdr}) into Eq.~(\ref{Eq:SWLPT-Adiab H_I,eff}) we find 
\begin{align}
\begin{split}
&\HO_{I,\text{eff}}^{\text{ad}}(t)= + 2\chi_{ac}|\eta_c(t)|^2 \hat{a}_l^{\dag} \hat{a}_l-2\chi_{ac}|\eta_c(t)|^2 \hat{a}_r^{\dag} \hat{a}_r \\
&-\frac{4\chi_{ac}^2|\eta_c(t)|^2(\Delta_{cd}+2\chi_{ac}\hat{a}_l^{\dag}\hat{a}_l+i\kappa_c/2)}{(\Delta_{cd}+2\chi_{ac}\hat{a}_l^{\dag}\hat{a}_l)^2+(\kappa_c/2)^2}\left(\hat{a}_l^{\dag}\hat{a}_l\right)^2\\
&+\frac{4\chi_{ac}^2|\eta_c(t)|^2(\Delta_{cd}+2\chi_{ac}\hat{a}_r^{\dag}\hat{a}_r-i\kappa_c/2)}{(\Delta_{cd}+2\chi_{ac}\hat{a}_r^{\dag}\hat{a}_r)^2+(\kappa_c/2)^2}\left(\hat{a}_r^{\dag}\hat{a}_r\right)^2\\
&+i \color{black}\frac{4\chi_{ac}^2 \kappa_c |\eta_c(t)|^2(\Delta_{cd}+2\chi_{ac}\hat{a}_l^{\dag}\hat{a}_l+i\kappa_c/2)}{[(\Delta_{cd}+2\chi_{ac}\hat{a}_l^{\dag}\hat{a}_l)^2+(\kappa_c/2)^2]} \\
& \times \frac{(\Delta_{cd}+2\chi_{ac}\hat{a}_r^{\dag}\hat{a}_r-i\kappa_c/2)}{[(\Delta_{cd}+2\chi_{ac}\hat{a}_r^{\dag}\hat{a}_r)^2+(\kappa_c/2)^2]}(\hat{a}_l^{\dag}\hat{a}_l)(\hat{a}_r^{\dag}\hat{a}_r) \;.
\end{split}
\label{Eq:SWLPT-Adiab H_I,eff expanded}
\end{align}

Equations~(\ref{Eq:SWLPT-Adiab H_I,eff})--(\ref{Eq:SWLPT-Adiab H_I,eff expanded}) are the main results of this Appendix. The real and imaginary parts of $\HO_{I,\text{eff}}^{\text{ad}}(t)$ provide the shift in qubit frequency and its dephasing rate due to the readout drive. 
%%%%%%%%%%%%%%%%%%%%%%%%%%%%%%%%%%%%%%%%%%%%%%%%%%%%%%%%%%%%%%%%%%%%%%%%%%%%%%%%

%%%%%%%%%%%%%%%%%%%%% App:Effective Map Prop %%%%%%%%%%%%%%%%%%%%%%%%%%%%%%%
\section{Properties of the effective dispersive map in the adiabatic regime}
\label{App:EffMapProp}

The effective adiabatic dispersive map for the vectorized density matrix is given by
\begin{align}
\ket{\Psi_{\hat{\rho},\text{eff}}^{\text{ad}}(t)} = e^{-i \int_{0}^{t} dt' \HO_{I,\text{eff}}^{\text{ad}}(t')} \ket{\Psi_{\hat{\rho},\text{eff}}(0)} \;,
\label{Eq:EffMapProp-Psi_eff(t) map}
\end{align}
where due to the diagonal form of $\HO_{I,\text{eff}}^{\text{ad}}(t)$ time-ordering is trivial and hence is dropped. In this appendix, we analyze the effective adiabatic dispersive spectrum, from which we prove desirable properties for the map such as TP and HP.  Moreover, in Appendix~\ref{App:LindForm}, we show that the map can be expressed in a Lindblad form, and hence is also CP.

\subsection{Effective adiabatic dispersive spectrum}
\label{SubApp:EffDispSpec}   

We define the effective adiabatic dispersive spectrum as eigenvalues of $\HO_{I,\text{eff}}^{\text{ad}}(t)$ in Eq.~(\ref{Eq:SWLPT-Adiab H_I,eff expanded}):
\begin{align}
\HO_{I,\text{eff}}^{\text{ad}}(t) \ket{n_{al}}\ket{n_{ar}} = E_{n_{al},n_{ar}}^{\text{ad}}(t) \ket{n_{al}}\ket{n_{ar}} \;.
\label{Eq:EffMapProp-Def of E_n_al,n_ar}
\end{align}
Given the diagonal form of Eq.~(\ref{Eq:SWLPT-Adiab H_I,eff expanded}), we find $ E_{n_{al},n_{ar}}^{\text{ad}}(t)$ immediately as 
\begin{widetext}
\begin{subequations}
\begin{align}
& \Re\{E_{n_{al},n_{ar}}^{\text{ad}}(t)\} = \frac{2 \chi_{ac} \left[\Delta_{cd}^2+(\frac{\kappa_c}{2})^2\right] \left[\left(\Delta_{cd} +2\chi_{ac}  n_{al}\right) \left(\Delta_{cd} +2 \chi_{ac} n_{ar}\right)+\left(\frac{\kappa_c}{2}\right)^2\right]\left(n_{al}-n_{ar}\right)|\eta_c(t)|^2}{\left[\left(\Delta_{cd} +2 \chi_{ac} n_{al}\right)^2+(\frac{\kappa_c}{2})^2\right] \left[\left(\Delta_{cd} +2 \chi_{ac}  n_{ar}\right)^2+(\frac{\kappa_c}{2})^2\right]} \;,
\label{Eq:EffMapProp-Re(E_nal,nar) Sol}\\
&\Im\{E_{n_{al},n_{ar}}^{\text{ad}}(t)\} = -\frac{2 \chi_{ac}^2 \kappa_c \left[\Delta_{cd}^2+(\frac{\kappa_c}{2})^2\right] \left(n_{al}-n_{ar}\right)^2 |\eta_c(t)|^2}{\left[\left(\Delta_{cd} +2 \chi_{ac} n_{al}\right)^2+(\frac{\kappa_c}{2})^2\right] \left[\left(\Delta_{cd} +2 \chi_{ac}  n_{ar}\right)^2+(\frac{\kappa_c}{2})^2\right]} \;, 
\label{Eq:EffMapProp-Im(E_nal,nar) Sol}
\end{align}
\end{subequations}
\end{widetext}
where the real and imaginary parts give the shift in the transition frequency and the dephasing of state $\ket{n_{al}}\ket{n_{ar}}$. Setting $n_{al}=1$ and $n_{ar}=0$ recovers $\Delta_{S}$ and $\gamma_{\phi}$ in Eqs.~(\ref{eqn:MeasIndStSh-Delta_S Sol}) and (\ref{eqn:MeasIndStSh-gamma_phi Sol}) of the main text.

\begin{figure}[t!]
\centering
\includegraphics[scale=0.44]{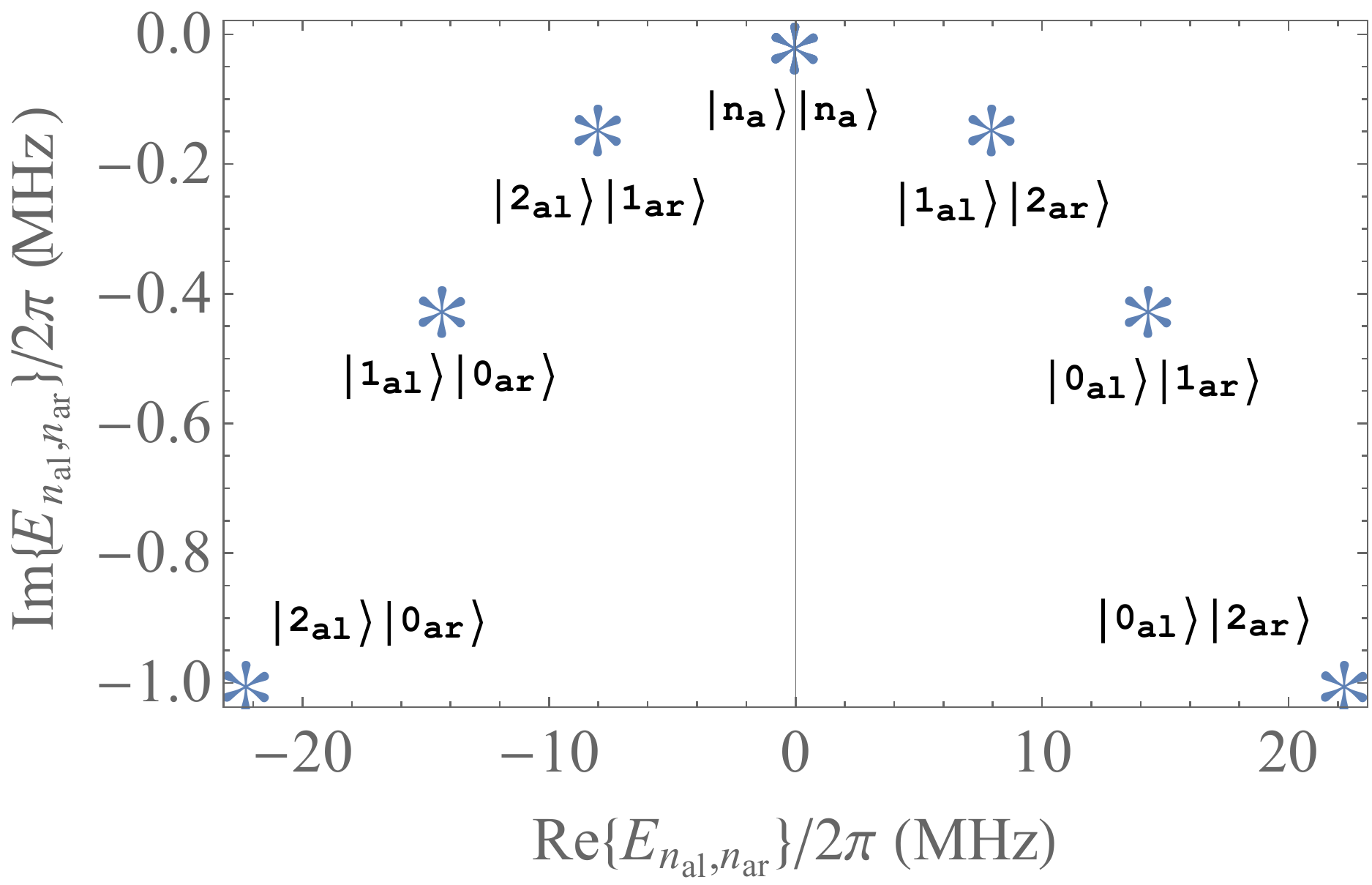}
\caption{Effective adiabatic dispersive spectrum in Eqs.~(\ref{Eq:EffMapProp-Re(E_nal,nar) Sol})--(\ref{Eq:EffMapProp-Im(E_nal,nar) Sol}) corresponding to $n_{al}$, $n_{ar}$ $\in \{0,1,2\}$. System parameters are set to $\Delta_{cd}/2\pi=-5$, $\chi_{ac}/2\pi=-1$, $\kappa_c/2\pi=1$ MHz, and $|\eta_c|^2=10$. Spectral properties~(\ref{Eq:EffMapProp-E_nn=0})--(\ref{Eq:EffMapProp-Im(E_mn)<0}) are clearly visible.}
\label{Fig:EffDispSpectrum}
\end{figure}

Based on Eqs.~(\ref{Eq:EffMapProp-Re(E_nal,nar) Sol})--(\ref{Eq:EffMapProp-Im(E_nal,nar) Sol}), the spectrum obeys the conditions: 
\begin{subequations}
\begin{align}
& E_{n_{a},n_{a}}^{\text{ad}}(t) = 0 \;,
\label{Eq:EffMapProp-E_nn=0}\\
& E_{n_{al},n_{ar}}^{\text{ad}}(t) = - E_{n_{ar},n_{al}}^{\text{ad}*}(t) \;,
\label{Eq:EffMapProp-E_mn=-E_nm^*}\\
& \Im\{E_{n_{al},n_{ar}}^{\text{ad}}(t)\}<0 \;, 
\label{Eq:EffMapProp-Im(E_mn)<0}
\end{align}
\end{subequations}
using which we prove some desirable properties for the map in the following subsections. Figure~\ref{Fig:EffDispSpectrum} shows the low-excitation spectrum based on Eqs.~(\ref{Eq:EffMapProp-Re(E_nal,nar) Sol})--(\ref{Eq:EffMapProp-Im(E_nal,nar) Sol}) and for a constant time-independent photon number.  

\subsection{Effective dispersive time evolution} 
\label{SubApp:EffDispEvol}
Given an arbitrary initial effective density matrix, we can express it in the vectorized form as
\begin{align}
\ket{\Psi_{\hat{\rho},\text{eff}}^{\text{ad}}(0)}=\sum\limits_{n_{al},n_{ar}} \rho_{n_{al},n_{ar}}^{\text{ad}} \ket{n_{al}}\ket{n_{ar}} \;,
\label{Eq:EffMapProp-Psi_rho expansion}
\end{align}
where $\rho_{n_{al},n_{ar}}^{\text{ad}}$ are the matrix elements. Based on the diagonal form of $\HO_{I,\text{eff}}^{\text{ad}}(t)$, the solution at time $t$ reads
\begin{align}
\ket{\Psi_{\hat{\rho},\text{eff}}^{\text{ad}}(t)}=\sum\limits_{n_{al},n_{ar}} \rho_{n_{al},n_{ar}}^{\text{ad}} e^{-i \int_{0}^{t} dt' E_{n_{al},n_{ar}}^{\text{ad}}(t')} \ket{n_{al}}\ket{n_{ar}} \;.
\label{Eq:EffMapProp-Psi_rho(t) Sol}
\end{align}
Undoing the vectorization, using Eqs.~(\ref{Eq:Vector-Rho expansion})--(\ref{Eq:Vector-corresponding Psi_rho}), the solution for the density matrix is found as
\begin{align}
\hat{\rho}_{\text{eff}}^{\text{ad}}(t)=\sum\limits_{n_{al},n_{ar}} \rho_{n_{al},n_{ar}}^{\text{ad}} e^{-i \int_{0}^{t} dt' E_{n_{al},n_{ar}}^{\text{ad}}(t')} \ket{n_{al}}\bra{n_{ar}} \;.
\label{Eq:EffMapProp-rho(t) Sol}
\end{align}

The time-evolution phase factors in Eq.~(\ref{Eq:EffMapProp-rho(t) Sol}) can be expressed as 
\begin{align}
\begin{split}
e^{-i \int_{0}^{t} dt' E_{n_{al},n_{ar}}^{\text{ad}}(t')} &= e^{-i \int_{0}^{t} dt' \Re\{E_{n_{al},n_{ar}}^{\text{ad}}(t')\}} \\
&\times e^{+ \int_{0}^{t} dt' \Im\{E_{n_{al},n_{ar}}^{\text{ad}}(t')\}} \;.
\end{split}
\label{Eq:EffMapProp-ConProp}
\end{align}
Based on Eq.~(\ref{Eq:EffMapProp-Im(E_nal,nar) Sol}), where $\Im\{E_{n_{al},n_{ar}}^{\text{ad}}(t)\}<0$, we find that individual matrix elements of $\hat{\rho}_{\text{eff}}^{\text{ad}}(t)$ decay in time and hence the map is contracting.

%%%%%%%%%%%%%%%%%%%%% SubApp: TP property %%%%%%%%%%%%%%%%%%%%%%%%%%%%%%%%%%%%
\subsection{Trace preservation}
\label{SubApp:TPProp}
To prove trace preservation, we need to show
\begin{align}
\sum\limits_{n_{a}} \rho_{n_{a},n_{a}}^{\text{ad}} e^{-i \int_{0}^{t} dt' E_{n_{a},n_{a}}^{\text{ad}}(t')} = \sum\limits_{n_{a}} \rho_{n_{a},n_{a}}^{\text{ad}} \;.
\label{Eq:EffMapProp-Def of EffMapProp}
\end{align}
However, we found in Eq.~(\ref{Eq:EffMapProp-E_nn=0}) that $E_{n_{a},n_{a}}^{\text{ad}}(t)=0$. This means that not only is the trace is preserved, as defined in Eq.~(\ref{Eq:EffMapProp-Def of EffMapProp}), but also individual diagonal elements of the density matrix $\rho_{n_{a},n_{a}}^{\text{ad}}$ are conserved.    
%%%%%%%%%%%%%%%%%%%%%%%%%%%%%%%%%%%%%%%%%%%%%%%%%%%%%%%%%%%%%%%%%%%%%%%%%%%%%%%

%%%%%%%%%%%%%%%% SubApp: HP Property %%%%%%%%%%%%%%%%%%%%%%%%%%%%%%%%%%%%%%%%%%
\subsection{Hermiticity preservation}
\label{SubApp:HPProp}

Taking the Hermitian conjugate of Eq.~(\ref{Eq:EffMapProp-rho(t) Sol}) we find
\begin{align}
\hat{\rho}_{\text{eff}}^{\text{ad}\dag}(t)=\sum\limits_{n_{al},n_{ar}} \rho_{n_{al},n_{ar}}^{\text{ad}*} e^{+i \int_{0}^{t} dt' E_{n_{al},n_{ar}}^{\text{ad}*}(t')} \ket{n_{ar}}\bra{n_{al}} \;.
\label{Eq:EffMapProp-rho(t)^dag}
\end{align}
Given that the initial density matrix is Hermitian, i.e.
\begin{align}
\rho_{n_{al},n_{ar}}^{\text{ad}*} = \rho_{n_{ar},n_{al}}^{\text{ad}} \;,
\end{align}
and using Eq.~(\ref{Eq:EffMapProp-E_mn=-E_nm^*}), we rewrite $\hat{\rho}_{\text{eff}}^{\text{ad}\dag}(t)$ as 
\begin{align}
\hat{\rho}_{\text{eff}}^{\text{ad}\dag}(t)=\sum\limits_{n_{al},n_{ar}} \rho_{n_{ar},n_{al}}^{\text{ad}} e^{-i \int_{0}^{t} dt' E_{n_{ar},n_{al}}^{\text{ad}}(t')} \ket{n_{ar}}\bra{n_{al}} \;.
\label{Eq:EffMapProp-rho(t)^dag 2}
\end{align}
Swapping the dummy indices $n_{al}$ and $n_{ar}$ in Eq.~(\ref{Eq:EffMapProp-rho(t)^dag 2}), we find that $\hat{\rho}_{\text{eff}}^{\text{ad}\dag}(t)=\hat{\rho}_{\text{eff}}^{\text{ad}}(t)$ and hence the map is Hermiticity preserving.
%%%%%%%%%%%%%%%%%%%%%%%%%%%%%%%%%%%%%%%%%%%%%%%%%%%%%%%%%%%%%%%%%%%%%%%%%%%%%%%

%%%%%%%%%%%%%%%%%%%% Fig: Effective Dispersive Spectrum%%%%%%%%%%%%%%%%%%%%%%%
%%%%%%%%%%%%%%%%%%%%%%%%%%%%%%%%%%%%%%%%%%%%%%%%%%%%%%%%%%%%%%%%%%%%%%%%%%%%%

%%%%%%%%%%%%%%%%%%%%%%%%%% App: LindForm %%%%%%%%%%%%%%%%%%%%%%%%%%%%%%%%%%%%%%%%
\section{Lindblad form of the effective dispersive map in the adiabatic regime}
\label{App:LindForm}

In this appendix, starting from Eq.~(\ref{Eq:SWLPT-Adiab H_I,eff}), we show that the effective adiabatic dispersive evolution can be rewritten in terms of an effective Lindblad evolution. To this aim, we group the terms in Eq.~(\ref{Eq:SWLPT-Adiab H_I,eff}) as follows. The first line and the real part of the second line provides the effective Hamiltonian or the coherent evolution, while the imaginary part of the second plus the third line contains the incoherent evolution.  

The incoherent part in Eq.~(\ref{Eq:SWLPT-Adiab H_I,eff}) can be expressed as
\begin{align}
\begin{split}
-\frac{i}{2} 4\chi_{ac}^2 \kappa_c |\eta_c(t)|^2\frac{\hat{n}_{al}^{2}}{|\hat{\Delta}_{cdl}|^2}\\
-\frac{i}{2} 4\chi_{ac}^2 \kappa_c |\eta_c(t)|^2\frac{\hat{n}_{ar}^{2}}{|\hat{\Delta}_{cdr}|^2}\\
+i 4\chi_{ac}^2 \kappa_c |\eta_c(t)|^2 \frac{\hat{n}_{al}}{\hat{\Delta}_{cdl}} \frac{\hat{n}_{ar}}{\hat{\Delta}_{cdr}} \;.
\end{split}
\label{Eq:LindForm-Incoh Part}
\end{align} 
To match Eq.~(\ref{Eq:LindForm-Incoh Part}) to a Lindblad dissipator of the form $i[\CO_{al} \CO_{ar}-(1/2)\CO_{al}^{\dag}\CO_{al}-(1/2)\CO_{ar}^{\dag}\CO_{ar}]$ (in the vectorized notation), we can define the following collapse operators:
\begin{subequations}
\begin{align}
&\CO_{al} \equiv \sqrt{4\chi_{ac}^2 \kappa_c |\eta_c(t)|^2} \frac{\hat{n}_{al}}{\hat{\Delta}_{cdl}} \;,
\label{Eq:LindForm-Def of C_al}\\
&\CO_{ar} \equiv \sqrt{4\chi_{ac}^2 \kappa_c |\eta_c(t)|^2} \frac{\hat{n}_{ar}}{\hat{\Delta}_{cdr}} \;.
\label{Eq:LindForm-Def of C_ar}
\end{align}
\end{subequations}

Reverting the vectorization, using Eqs.~(\ref{Eq:Vector-Def of MathcalO_l})--(\ref{Eq:Vector-Def of MathcalO_r}), we arrive at the following effective Lindblad dynamics:
\begin{align}
\dot{\hat{\rho}}_{\text{I,eff}}^{\text{ad}}(t)=-i[\hat{H}_{\text{I,eff}}^{\text{ad}}(t),\hat{\rho}_{\text{I,eff}}^{\text{ad}}(t)]+\mathcal{D}[\hat{C}_{\text{I,eff}}^{\text{ad}}(t)]\hat{\rho}_{\text{I,eff}}^{\text{ad}}(t) \;.
\label{Eq:LindForm-Eff Lind dynamics}
\end{align}
The coherent part is generated by a Hamiltonian containing the first- and second-order Stark shifts
\begin{align}
\begin{split}
\hat{H}_{\text{I,eff}}^{\text{ad}}&=2\chi_{ac}|\eta_c(t)|^2 \hat{n}_{a} \\
&-\frac{4\chi_{ac}^2|\eta_c(t)|^2(\Delta_{cd}+2\chi_{ac}\hat{n}_{a})}{(\Delta_{cd}+2\chi_{ac}\hat{n}_{a})^2+(\kappa_c/2)^2}\hat{n}_{a}^2 \;,
\label{Eq:LindForm-Def of H_I,eff^ad}
\end{split}
\end{align}
while the incoherent part accounts for measurement-induced dephasing in terms of the effective collapse operator  
\begin{align}
\begin{split}
\hat{C}_{\text{I,eff}}^{\text{ad}}(t) = \frac{\sqrt{4\chi_{ac}^2\kappa_c|\eta_c(t)|^2}\hat{n}_{a}}{\Delta_{cd}-i\kappa_c/2+2\chi_{ac}\hat{n}_a} \;,
\end{split}
\label{Eq:LindForm-Def of C_I,eff^ad}
\end{align}
for which we have that $\CO_{al} = \hat{C}_{\text{I,eff}}^{\text{ad}}(t) \otimes \hat{I}$ and $\CO_{ar} = \hat{I} \otimes \hat{C}_{\text{I,eff}}^{\text{ad} *}(t)$ as required.
Given the Lindblad form, we conclude that the effective adiabatic dispersive map is CPTP.
%%%%%%%%%%%%%%%%%%%%%%%%%%%%%%%%%%%%%%%%%%%%%%%%%%%%%%%%%%%%%%%%%%%%%%%%%%%%

%%%%%%%%%%%%%%%%%%%%%%%%%%%%%% App: Transient behavior %%%%%%%%%%%%%%%%%%%%%%%%%%
\section{Transient behavior of effective interactions}
\label{App:Transient}

In this appendix, starting from the time-dependent form of the effective extended Hamiltonian, we analyze the transient behavior and dependence on the pulse shape. 

Putting the time-dependent contributions of Appendix~\ref{App:SWLPT} together we find $\HO_{I,\text{eff}}(t)$ as
\begin{align}
\begin{split}
\HO_{I,\text{eff}}(t) &=  2\chi_{ac}|\eta_c(t)|^2 \hat{n}_{al}-2\chi_{ac}|\eta_c(t)|^2 \hat{n}_{ar}\\
&-4\chi_{ac}^2\AO_{\eta,ll}(t)\hat{n}_{al}^2+4\chi_{ac}^2\AO_{\eta,rr}(t)\hat{n}_{ar}^2 \\
&+i4\chi_{ac}^2\kappa_c \bigg[\frac{\hat{\mathcal{B}}_{\eta,lr}(t)}{6}+\frac{\hat{\mathcal{C}}_{\eta,lr}(t)}{2}\bigg]\hat{n}_{al}\hat{n}_{ar} \;,
\end{split}
\label{Eq:Transient-TimeDep H_I,eff(t)}
\end{align}
with $\AO_{\eta,ll}(t)$, $\AO_{\eta,rr}(t)$, $\BO_{\eta,lr}(t)$ and $\CO_{\eta,lr}(t)$ given in Eqs.~(\ref{Eq:SWLPT-Def of A_eta,ll}), (\ref{Eq:SWLPT-Def of A_eta,rr}), (\ref{Eq:SWLPT-Def of B_eta,lr}) and (\ref{Eq:SWLPT-Def of C_eta,lr}), respectively.

We note that we consider three ways to compute the time-dependent correlation functions: (i) integrals in time-domain, as found by the solutions to the SWLPT ODEs [Eqs.~(\ref{Eq:SWLPT-Def of A_eta,ll}), (\ref{Eq:SWLPT-Def of A_eta,rr}), (\ref{Eq:SWLPT-Def of B_eta,lr}) and (\ref{Eq:SWLPT-Def of C_eta,lr})], (ii) adiabatic expansion, which brings corrections in terms of the pulse shape derivatives, and (iii) Fourier representation, which connects the response directly to the Fourier transform of the input pulse. The first step is to obtain the solution for the resonator response $\eta_c(t)$, based on Eq.~(\ref{Eq:DispTrans-Cond for eta_c(t)}), either analytically or numerically. In the following, we discuss the adiabatic expansion and Fourier representation of the correlation functions.    

\subsection{Adiabatic expansion}
\label{SubApp:TransAdiabExp}

Consider the second-order correlation $\AO_{\eta,jj}(t)$ as 
\begin{align}
\begin{split}
\hat{\mathcal{A}}_{\eta,jj}(t) & \equiv \frac{1}{2i}\int^{t}dt' \eta_c(t)\eta_c^*(t')e^{i\hat{\Delta}_{cdj}(t-t')} \\
&-\frac{1}{2i}\int^{t}dt' \eta_c^*(t)\eta_c(t')e^{-i\hat{\Delta}_{cdj}(t-t')} \;,
\end{split}
\label{Eq:Transient-Def of A_eta,jj}
\end{align}
with $j \in \{l, r\}$. The idea for an adiabatic expansion is to compute the integrals via integration by parts, which leads to a series expansion in terms of the derivatives of $\eta_c(t)$ and $\eta_c^*(t)$.  

In particular, the integral in the first line of Eq.~(\ref{Eq:Transient-Def of A_eta,jj}) can be expressed as
\begin{align}
\int^{t}dt'\eta_c^*(t') e^{-i\hat{\Delta}_{cdj}t'} =\sum\limits_{n=0}^{\infty} \frac{(-1)^n \frac{d^n \eta_c^*(t)}{dt^n}}{(-i\hat{\Delta}_{cdj})^{n+1}} e^{-i\hat{\Delta}_{cdj}t} \;.
\label{Eq:Transient-IntegralByParts}
\end{align}
In finding Eq.~(\ref{Eq:Transient-IntegralByParts}), we use integration by parts: 
\begin{align}
\begin{split}
&\int^{t}dt'\eta_c^*(t') e^{-i\hat{\Delta}_{cdj}t'} =\\
&\int^{t}dt'\eta_c^*(t') \left(\frac{1}{-i\hat{\Delta}_{cdj}}\frac{d}{dt'}\right)e^{-i\hat{\Delta}_{cdj}t'}=\\
&\frac{\eta_c^*(t)}{-i\hat{\Delta}_{cdj}}e^{-i\hat{\Delta}_{cdj}t}-\int^{t}dt'\frac{1}{-i\hat{\Delta}_{cdj}}\frac{d\eta_c^*(t')}{dt'} e^{-i\hat{\Delta}_{cdj}t'} \;,
\end{split}
\label{Eq:Transient-IntegralByParts1storder}
\end{align}
and repeat to infinite order.

Using Eq.~(\ref{Eq:Transient-IntegralByParts}), and similarly for $\int^{t}dt'\eta_c(t') e^{+i\hat{\Delta}_{cdj}t'}$, we express $\AO_{\eta,jj}(t)$ as
\begin{align}
\begin{split}
\AO_{\eta,jj}(t)&=\frac{1}{2i}\sum\limits_{n=0}^{\infty} \frac{(-1)^n \eta_c(t)\frac{d^n \eta_c^*(t)}{dt^n}}{(-i\hat{\Delta}_{cdj})^{n+1}} \\
&-\frac{1}{2i}\sum\limits_{n=0}^{\infty} \frac{(-1)^n \eta_c^*(t)\frac{d^n \eta_c(t)}{dt^n}}{(+i\hat{\Delta}_{cdj})^{n+1}} \;.
\end{split}
\label{Eq:Transient-AdiabExp of A_eta,jj}
\end{align}
The first few terms in the adiabatic expansion~(\ref{Eq:Transient-AdiabExp of A_eta,jj}) are
\begin{align}
\begin{split}
\AO_{\eta,jj}(t)&=\frac{|\eta_c(t)|^2}{\hat{\Delta}_{cdj}}+\frac{\eta_c(t)\dot{\eta}_c^{*}(t)-\eta_c^*(t)\dot{\eta}_c(t)}{2i \hat{\Delta}_{cdj}^2}\\
&-\frac{\eta_c(t)\ddot{\eta}_c^{*}(t)+\eta_c^*(t)\ddot{\eta}_c(t)}{2 \hat{\Delta}_{cdj}^3}+O\left(\left|\frac{\eta_c(t)\dddot{\eta}_c^*(t)}{\hat{\Delta}_{cdj}^4}\right|\right) \;.
\end{split}
\end{align}
The first and the second terms are referred to as the dynamic and geometric contributions~\cite{Cross_Optimized_2015}. 

The accuracy of the adiabatic expansion depends on the effective gap $|\braket{\hat{\Delta}_{cdj}}|$, which in turn depends on $\Delta_{cd}$, $2\chi_{ac}$ and $\kappa_c$, and their relation to the time scale of the pulse ramp. Similar adiabatic expansions can be derived for $\BO_{\eta,lr}(t)$ and $\CO_{\eta,lr}(t)$.     

\subsection{Fourier representation}
\label{SubApp:TransFourier}

An alternative representation of the correlation functions can be found in the frequency domain. We define the Fourier and the inverse Fourier transforms as:
\begin{subequations}
\begin{align}
&\tilde{f}(\omega) \equiv \int_{-\infty}^{\infty} dt\; f(t) e^{-i\omega t} \;,
\label{Eq:Transient-Def of Fourier}\\
&f(t) = \int_{-\infty}^{\infty} \frac{d\omega}{2\pi}\;  \tilde{f}(\omega) e^{i\omega t} \;.
\label{Eq:Transient-Def of InvFourier}
\end{align}
\end{subequations}
Using Eqs.~(\ref{Eq:Transient-Def of Fourier})--(\ref{Eq:Transient-Def of InvFourier}), $\AO_{\eta,jj}(t)$ can be written as
\begin{align}
\begin{split}
&\AO_{\eta,jj}(t)=\int_{-\infty}^{\infty} \frac{d\omega}{2\pi}\int_{-\infty}^{\infty}\frac{d\omega'}{2\pi} \\
&\bigg[\frac{(\omega+\omega'+2\hat{\Delta}_{cdj})\tilde{\eta}_c^*(\omega)\tilde{\eta}_c(\omega')}{2(\omega+\hat{\Delta}_{cdj})(\omega'+\hat{\Delta}_{cdj})}e^{-i(\omega-\omega')t} \bigg]\;,
\label{Eq:Transient-FourierRep of A_eta,jj}
\end{split}	
\end{align}
for $j\in\{l, r\}$. In writing Eq.~(\ref{Eq:Transient-FourierRep of A_eta,jj}), we grouped and simplified the contributions in terms of a common Fourier basis $\exp[-i(\omega-\omega')t]$. Similarly, the Fourier representation of $\BO_{\eta,lr}(t)$ reads  
\begin{align}
\begin{split}
&\BO_{\eta,lr}(t)=\int_{-\infty}^{\infty} \frac{d\omega}{2\pi}\int_{-\infty}^{\infty}\frac{d\omega'}{2\pi} \\
&\bigg\{\frac{[\omega'-\omega+3(\hat{\Delta}_{cdl}-\hat{\Delta}_{cdr})]\tilde{\eta}_c^*(\omega)\tilde{\eta}_c(\omega')}{2(\omega+\hat{\Delta}_{cdl})(\omega'+\hat{\Delta}_{cdr})(\hat{\Delta}_{cdl}-\hat{\Delta}_{cdr})}e^{-i(\omega-\omega')t} \bigg\}\;.
\end{split}
\label{Eq:Transient-FourierRep of B_eta,lr}
\end{align}
For $\CO_{\eta,lr}(t)$ we find
\begin{align}
\begin{split}
&\CO_{\eta,lr}(t)=\int_{-\infty}^{\infty} \frac{d\omega}{2\pi}\int_{-\infty}^{\infty}\frac{d\omega'}{2\pi} \\
&\bigg\{\frac{(\hat{\Delta}_{cdl}-2\hat{\Delta}_{cdr}-\omega)\tilde{\eta}_c^*(\omega)\tilde{\eta}_c(\omega')}{2(\omega+\hat{\Delta}_{cdl})(\omega+\hat{\Delta}_{cdr})(\hat{\Delta}_{cdl}-\hat{\Delta}_{cdr})}e^{-i(\omega-\omega')t} \\
&+ \frac{(2\hat{\Delta}_{cdl}-\hat{\Delta}_{cdr}+\omega')\tilde{\eta}_c^*(\omega)\tilde{\eta}_c(\omega')}{2(\omega'+\hat{\Delta}_{cdl})(\omega'+\hat{\Delta}_{cdr})(\hat{\Delta}_{cdl}-\hat{\Delta}_{cdr})}e^{-i(\omega-\omega')t}\bigg\}\;.
\end{split}
\label{Eq:Transient-FourierRep of C_eta,lr}
\end{align}

We note that, based on Eq.~(\ref{Eq:DispTrans-Cond for eta_c(t)}), the resonator response $\eta_c(t)$ has the following explicit solution in the Fourier domain   
\begin{align}
\tilde{\eta}_c(\omega)=\frac{\tilde{\Omega}_c(\omega)}{2(\omega-\Delta_{cd0})} \;,
\label{Eq:Transient-eta_c(w) Sol}
\end{align}
where $\tilde{\Omega}_c(\omega)$ is the Fourier transform of the input pulse and $\Delta_{cd0}\equiv \Delta_{cd}-i\kappa_c/2$. Therefore, the Fourier representations~(\ref{Eq:Transient-FourierRep of A_eta,jj})--(\ref{Eq:Transient-FourierRep of C_eta,lr}) can also be directly expressed in terms of the Fourier transform of the drive pulse $\tilde{\Omega}_c(\omega)$.

%%%%%%%%%%%%%%%%%%%%%%%%%%%%%%%%%%%%%%%%%%%%%%%%%%%%%%%%%%%%%%%%%%%%%%%%%%%%%%%%%

%%%%%%%%%%%%%%%%%%%%%%%%% App: Instantaneous eigenstates %%%%%%%%%%%%%%%%%%%%%%%
\section{Instantaneous eigenstates}
\label{App:EigState}
Following the methods introduced in Appendices~\ref{App:Vector},~\ref{App:DispTrans} and~\ref{App:SWLPT}, we have computed an effective diagonal generator for the evolution. In particular, in this effective frame, states of the resonator are integrated out. Here, we provide the representation of the corresponding eigenstates in the starting frame, i.e. rotating frame of the drive, which makes the role of the resonator modes more explicit.

Note that our diagonalization employed three intermediate transformations: (i) displacement transformation of the resonator mode, (ii) transformation to the interaction frame, and (iii) SW transformation:
\begin{align}
\TO_{\text{diag}}(t) \equiv \TO_{D}[\eta_c(t)]
\TO_{0}(t) \TO_{\text{SW}}(t) \;,
\label{Eq:EigState-Def of Udiag(t)}
\end{align}
with $\TO_{D}[\eta_c(t)]$ given in Eq.~(\ref{Eq:DispTrans-Def of D[eta]}), $\TO_{0}(t)$ being the interaction frame with respect to Eqs.~(\ref{Eq:SWLPT-Def of H_l0})--(\ref{Eq:SWLPT-Def of H_r0}), and $\TO_{\text{SW}}(t)$ solved for perturbatively in Appendix~\ref{App:SWLPT}. 

Using Eq.~(\ref{Eq:EigState-Def of Udiag(t)}), we re-expressed the starting (rotating frame of the drive) vectorized Lindblad dynamics,
\begin{align}
\left[\HO_u(t)-i\partial_t\right] \ket{\Psi_{\hat{\rho}}(t)} = 0 \;,
\label{Eq:EigState-VectLindEq RotFrame}
\end{align}
in the effective frame as 
\begin{align}
\TO_{\text{diag}}^{-1}(t)\left[\HO_u(t)-i\partial_t\right] \TO_{\text{diag}}(t)\TO_{\text{diag}}^{-1}(t)\ket{\Psi_{\hat{\rho}}(t)} = 0 \;.
\label{Eq:EigState-VectLindEq EffFrame}
\end{align}
Therefore, the starting and effective vectorized density matrices are related via
\begin{align}
\ket{\Psi_{\hat{\rho}}(t)} = \TO_{\text{diag}}(t) \ket{\Psi_{\hat{\rho},\text{eff}}(t)} \;.
\end{align}
Consequently, the eigenstates of the extended Hamiltonian $\HO_u(t)$ are represented in the rotating frame of the drive as 
\begin{align}
\ket{\Psi_{n_{al},n_{cl},n_{ar},n_{cr}}(t)} \equiv \TO_{\text{diag}}(t) \ket{n_{al},n_{cl}}\ket{n_{ar},n_{cr}} \;,
\label{Eq:EigState-Def of Psi_alclarcr}
\end{align}
where the right hand side is the number basis in the effective (diagonal) frame.

The SW transformation $\TO_{\text{SW}}(t)$ can be computed perturbatively as 
\begin{align}
\begin{split}
\TO_{\text{SW}}(t) \equiv e^{-i\GO (t)} & = \IO - i\GO_1(t)\\
&-i\GO_2(t)-\frac{1}{2}\GO_1^2(t) \\
& + O\left(\HO_I^3(t)\right) \;,
\end{split}
\label{Eq:EigState-Def of Usw(t)}
\end{align}
with $\GO_1(t)$ and $\GO_2(t)$ given in Eqs.~(\ref{Eq:SWLPT-G1 Sol}) and (\ref{Eq:SWLPT-G2 Sol}), respectively. Using the lowest order adiabatic expansion, under which $\HO_{I,\text{eff}}^{\text{ad}}(t)$ in Eq.~(\ref{Eq:SWLPT-Adiab H_I,eff expanded}) was derived, we find $\GO_1^{\text{ad}}(t)$ as
\begin{align}
\begin{split}
\GO_{1}^{\text{ad}} (t) & = i \frac{2\chi_{ac} \eta_c^*(t)}{\hat{\Delta}_{cdl}}e^{-i\hat{\Delta}_{cdl}t} \hat{a}_l^{\dag}\hat{a}_l \hat{c}_l \\
&- i \frac{2\chi_{ac} \eta_c(t)}{\hat{\Delta}_{cdl}} e^{i\hat{\Delta}_{cdl}t}\hat{a}_l^{\dag}\hat{a}_l \hat{c}_l^{\dag}\\
&+ i \frac{2\chi_{ac} \eta_c(t)}{\hat{\Delta}_{cdr}} e^{i\hat{\Delta}_{cdr}t} \hat{a}_r^{\dag}\hat{a}_r \hat{c}_r \\
&- i \frac{2\chi_{ac} \eta_c^*(t)}{\hat{\Delta}_{cdr}} e^{-i\hat{\Delta}_{cdr}t} \hat{a}_r^{\dag}\hat{a}_r \hat{c}_r^{\dag}\\
&+ \frac{\kappa_c}{\hat{\Delta}_{cdr}-\hat{\Delta}_{cdl}} e^{i(\hat{\Delta}_{cdr}-\hat{\Delta}_{cdl})t} \hat{c}_l \hat{c}_r  \;.  
\end{split}
\label{Eq:EigState-Ad G1 Sol}
\end{align}
and $\GO_2^{\text{ad}}(t)$ as
\begin{align}
\begin{split}
\GO_{2}^{\text{ad}} (t) & = \chi_{ac}\kappa_c \eta_c(t)\frac{2\hat{\Delta}_{cdl}-\hat{\Delta}_{cdr}}{\hat{\Delta}_{cdr}\hat{\Delta}_{cdl}(\hat{\Delta}_{cdr}-\hat{\Delta}_{cdl})}e^{i\hat{\Delta}_{cdr}t} \hat{a}_l^{\dag}\hat{a}_l \hat{c}_r \\
&-\chi_{ac}\kappa_c \eta_c^*(t)\frac{2\hat{\Delta}_{cdr}-\hat{\Delta}_{cdl}}{\hat{\Delta}_{cdr}\hat{\Delta}_{cdl}(\hat{\Delta}_{cdl}-\hat{\Delta}_{cdr})}e^{-i\hat{\Delta}_{cdl}t} \hat{a}_r^{\dag}\hat{a}_r \hat{c}_l \;.
\end{split}
\label{Eq:EigState-Ad G2 Sol}
\end{align}

As explicit examples, we provide expressions for the eigenstates that correspond to the computational subspace of the qubit in the rotating frame of the drive. Up to the zeroth order in SWLPT (keeping only $\IO$ in Eq.~(\ref{Eq:EigState-Def of Usw(t)})), the resonator degrees of freedom are in the coherent state characterized by $\eta_c(t)$. Higher-order processes, however, can create/annihilate resonator excitations depending on the qubit state according to Eqs.~(\ref{Eq:EigState-Ad G1 Sol})--(\ref{Eq:EigState-Ad G2 Sol}). Using Eq.~(\ref{Eq:EigState-Def of Psi_alclarcr}), and up to the second order in SWLPT, we find $\ket{\Psi_{0_{al},0_{cl},0_{ar},0_{cr}}^{(2),\text{ad}}(t)}$ as
\begin{subequations}
\begin{align}
\ket{\Psi_{0_{al},0_{cl},0_{ar},0_{cr}}^{(2),\text{ad}}(t)} = \ket{0_{al},\eta_c(t)}\ket{0_{ar},\eta_c^*(t)} \;,
\label{Eq:EigState-Psi_0l,0r Sol}
\end{align}
implying that when the qubit is in the ground state, the coherent resonator state remains an eigenstate of the extended dispersive Hamiltonian. For $\ket{\Psi_{1_{al},0_{cl},0_{ar},0_{cr}}^{(2),\text{ad}}(t)}$, however, we find
\begin{align}
\begin{split}
&\ket{\Psi_{1_{al},0_{cl},0_{ar},0_{cr}}^{(2),\text{ad}}(t)} = \ket{1_{al},\eta_c(t)}\ket{0_{ar},\eta_c^*(t)} \\
&-\frac{2\chi_{ac}\eta_c(t)\left[\hat{c}_l^{\dag}-\eta_c^*(t)\right]}{\Delta_{cd}-i\frac{\kappa_c}{2}+2\chi_{ac}}\ket{1_{al},\eta_c(t)}\ket{0_{ar},\eta_c^*(t)}\\
&+ \frac{2\chi_{ac}^2\eta_c^2(t)\left[\hat{c}_l^{\dag}-\eta_c^*(t)\right]^2}{[\Delta_{cd}-i\frac{\kappa_c}{2}+2\chi_{ac}]^2}\ket{1_{al},\eta_c(t)}\ket{0_{ar},\eta_c^*(t)},
\end{split}
\label{Eq:EigState-Psi_1l,0r Sol}
\end{align}
where the second and the third lines come from $-i\GO_1(t)$ and $-(1/2)\GO_1^2(t)$, respectively. Similarly, $\ket{\Psi_{0_{al},0_{cl},1_{ar},0_{cr}}^{(2),\text{ad}}(t)}$ reads
\begin{align}
\begin{split}
&\ket{\Psi_{0_{al},0_{cl},1_{ar},0_{cr}}^{(2),\text{ad}}(t)} = \ket{0_{al},\eta_c(t)}\ket{1_{ar},\eta_c^*(t)} \\
&-\frac{2\chi_{ac}\eta_c^*(t)\left[\hat{c}_r^{\dag}-\eta_c(t)\right]}{\Delta_{cd}+i\frac{\kappa_c}{2}+2\chi_{ac}}\ket{0_{al},\eta_c(t)}\ket{1_{ar},\eta_c^*(t)}\\
&+\frac{2\chi_{ac}^2\eta_c^{*2}(t)\left[\hat{c}_r^{\dag}-\eta_c(t)\right]^2}{[\Delta_{cd}+i\frac{\kappa_c}{2}+2\chi_{ac}]^2}\ket{0_{al},\eta_c(t)}\ket{1_{ar},\eta_c^*(t)}.
\end{split}
\label{Eq:EigState-Psi_0l,1r Sol}
\end{align}
Lastly, $\ket{\Psi_{1_{al},0_{cl},1_{ar},0_{cr}}^{(2),\text{ad}}(t)}$ is found as	
\begin{align}
\begin{split}
&\ket{\Psi_{1_{al},0_{cl},1_{ar},0_{cr}}^{(2),\text{ad}}(t)} = \ket{1_{al},\eta_c(t)}\ket{1_{ar},\eta_c^*(t)} \\
&-\frac{2\chi_{ac}\eta_c(t)\left[\hat{c}_l^{\dag}-\eta_c^*(t)\right]}{\Delta_{cd}-i\frac{\kappa_c}{2}+2\chi_{ac}}\ket{1_{al},\eta_c(t)}\ket{1_{ar},\eta_c^*(t)}\\
&-\frac{2\chi_{ac}\eta_c^*(t)\left[\hat{c}_r^{\dag}-\eta_c(t)\right]}{\Delta_{cd}+i\frac{\kappa_c}{2}+2\chi_{ac}}\ket{1_{al},\eta_c(t)}\ket{1_{ar},\eta_c^*(t)} \\
&+ \frac{2\chi_{ac}^2\eta_c^2(t)\left[\hat{c}_l^{\dag}-\eta_c^*(t)\right]^2}{\left(\Delta_{cd}-i\frac{\kappa_c}{2}+2\chi_{ac}\right)^2}\ket{1_{al},\eta_c(t)}\ket{1_{ar},\eta_c^*(t)} \\
&+\frac{2\chi_{ac}^2\eta_c^{*2}(t)\left[\hat{c}_r^{\dag}-\eta_c(t)\right]^2}{\left(\Delta_{cd}+i\frac{\kappa_c}{2}+2\chi_{ac}\right)^2}\ket{1_{al},\eta_c(t)}\ket{1_{ar},\eta_c^*(t)}\\
&+\frac{4\chi_{ac}^2|\eta_c(t)|^2\left[\hat{c}_l^{\dag}-\eta_c^*(t)\right]\left[\hat{c}_r^{\dag}-\eta_c(t)\right]}{(\Delta_{cd}+2\chi_{ac})^2+(\frac{\kappa_c}{2})^2}\\
&\times\ket{1_{al},\eta_c(t)}\ket{1_{ar},\eta_c^*(t)} \;.
\end{split}
\end{align}
\label{Eq:EigState-Psi_1l,1r Sol}
\end{subequations}
%%%%%%%%%%%%%%%%%%%%%%%%%%%%%%%%%%%%%%%%%%%%%%%%%%%%%%%%%%%%%%%%%%%%%%%%%%%%%%%%%

%%%%%%%%%%%%%%%%%%%%%%%%%%%%%%%%%%%%%%%%%%%%%%%%%%%%%%%%%%%%%%%%%%%%%%%%%%%%%%%%%
\bibliographystyle{unsrt}
\bibliography{SWPTForReadoutBibliography}
\end{document}